\documentclass[10pt]{article}

\usepackage{amsmath}
\usepackage{amssymb}
\usepackage{graphicx}

\usepackage{cite}

\usepackage{color} 

\usepackage{multirow}
\usepackage{lscape}

\usepackage[labelfont=bf,labelsep=period,justification=raggedright]{caption}
\usepackage{pstricks,pst-grad,pst-plot}

\makeatletter
\renewcommand{\@biblabel}[1]{\quad#1.}
\makeatother

\date{}

\begin{document}

\begin{flushleft}
{\Large
\textbf{Prediction of invasion from the early stage of an epidemic}
}
\\
Francisco~J.~P{\'e}rez-Reche$^{1,\ast}$, 
Franco~M.~Neri$^{2}$, 
Sergei~N.~Taraskin$^{3}$,
Christopher~A.~Gilligan$^{2}$
\\
\bf{1} SIMBIOS Centre, University of Abertay Dundee,  Dundee,  UK
\\
\bf{2} Department of Plant Sciences, University of Cambridge, Cambridge,  UK
\\
\bf{3} St. Catharine's College and Department of Chemistry, University of Cambridge, Cambridge, UK
\\
$\ast$ E-mail: p.perezreche@abertay.ac.uk
\end{flushleft}

\section*{Abstract}
Predictability of undesired events is a question of great interest in many scientific disciplines including
seismology, economy, and epidemiology. Here, we focus on the predictability of invasion of a broad
class of epidemics caused by diseases that lead to permanent immunity of infected hosts after recovery
or death. We approach the problem from the perspective of the science of complexity by proposing
and testing several strategies for the estimation of important characteristics of epidemics, such as the
probability of invasion. Our results suggest that parsimonious approximate methodologies may lead
to the most reliable and robust predictions. The proposed methodologies are first applied to analysis of
experimentally observed epidemics: invasion of the fungal plant pathogen \emph{Rhizoctonia solani} in replicated
host microcosms. We then consider numerical experiments of the SIR (susceptible-infected-removed)
model to investigate the performance of the proposed methods in further detail. The suggested framework
can be used as a valuable tool for quick assessment of epidemic threat at the stage when epidemics
only start developing. Moreover, our work amplifies the significance of the small-scale and finite-time
microcosm realizations of epidemics revealing their predictive power.

\section{Introduction}

Predictability of catastrophic events such as earthquakes, epidemics, fracture or financial crashes~\cite{Sornette2000,Sornette_PNAS2002,Sornette_StockCrash2003} is a topic of increasing interdisciplinary interest.  The predictability of these events is inextricably linked to the inherent complexity of the phenomena under consideration \cite{Sornette2000,Sornette_PNAS2002}. Here, we focus on epidemiology.
Within this context, many studies have been devoted to prediction of the
temporal incidence of epidemics (i.e. the evolution of the number of infected
hosts in the course of an epidemic)
\cite{Medley_Science2001,Valleron_Science2001_vCJDPrediction,HuillardAignaux_Science2001,Morton_JRoySocC2005,Kleczkowski_JRSocInterf2007}.
Recently, an increasing number of papers have also considered the prediction of the
spatio-temporal evolution of epidemics \cite{Keeling_Science2001,Ferguson_Science2001_FootMouthPred,Hufnagel_PNAS2004,Riley_Science2007}.

Both the temporal and the spatio-temporal incidence depend on complex
factors related to  the transmission of infection and the properties
of the hosts.
For instance, the hosts are not identical in susceptibility
and transmissibility of infection due to difference in age, size, genotype and
neighbourhood. \cite{Keeling_Science2001,Ferguson_Science2001_FootMouthPred,Valleron_Science2001_vCJDPrediction,PerezReche_JRSInterface2010}.
The transmission of infection is stochastic, meaning that a healthy host
is infected by contact with inoculum from an infected host with a certain
probability only. Many epidemics, notably those involving transmission by invertebrate vectors, or by wind and rain for many plant pathogens, are subject to variability in weather. This environmental stochasticity can also influence the
evolution of epidemics in such heterogeneous systems
\cite{Truscott_PNAS2003}.
All these factors make prediction of
disease incidence an extremely challenging and sometimes controversial
task~\cite{Medley_Science2001,Dye_Science2003_Perspective,May_Science2004}.
In  Ref.~\cite{Medley_Science2001},
it was suggested that, although obtaining precise quantitative predictions for
the incidence would be obviously desirable, qualitative predictions
may be more valuable.
This is very much along the
lines of ideas from the science of complexity claiming that, despite the fact
that giving accurate predictions for the detailed evolution of complex systems
might be an illusory task,
certain qualitative features of the evolution,
such as occurrence or absence of a catastrophic event,
 could be more amenable for
prediction~\cite{Sornette_PNAS2002,Pearce_IntJEpidemiol2006_ComplexEpidemics,Goldenfel_Science1999_Complexity}.

Here, we address the question of predictability of epidemics using a
methodological framework inspired by the science of complexity.  The
main aim is to estimate the probability that an emerging epidemic will
invade a significant fraction of the population in the future.  This
quantity can be viewed as a qualitative feature of the complete
spatio-temporal evolution of epidemics.

We propose several methods for approaching the problem that offer different levels of precision. Our results suggest that the most precise methods do not necessarily lead to more reliable predictions. Instead, parsimony seems to be the key ingredient for prediction based on inherently limited observations.
The framework  presented below deals with epidemics caused by a broad class of pathogens leading to
permanent immunity of infected hosts after
recovery (or death).
There are numerous examples of such
diseases affecting populations of
humans\cite{AndersonMay_Book1991,Murray_02:book},
animals\cite{Davis_Nature2008} and
plants\cite{Otten_Ecology2003_rates}.
The advantage in analysis of such epidemics
is that they are
characterised by a well-defined final state consisting of only hosts
that were never infected and hosts that were infected and became
immune.
In particular, we focus on the estimation of the probability that an
epidemic is invasive in the final state.

The proposed methods are  first applied to prediction of invasion of a pathogen in an experimental model system in which the fungal plant pathogen, \emph{Rhizoctonia solani}, spreads through a population of hosts represented by discrete  nutrient sites. The properties of the sites, e.g. nutrient concentration, can be varied for  different realisations of epidemics. Such a system is convenient for generation and observation of rapid, highly replicated and repeatable epidemics and it is used as a benchmark for description of our methodologies and analyses. 
The epidemic prediction analysis for experimental system is followed by a
test of our methods in 
numerical experiments for epidemics spreading on networks of hosts arranged on a regular lattice. The advantage of investigating such epidemics is that their properties are known beforehand and this allows us to provide a precise analysis of the performance of prediction  methods by comparing with the expected behaviour.

\section{Methods}
\label{Sec:Methods}
The methods follow four steps that are basic for any scientifically
meaningful prediction of the behaviour of a complex system:
(i) Observation of the initial evolution of the process over a certain
period of time, $t_{\text{obs}}$.
(ii) Construction of a model for
description of the observed behaviour.
(iii) Fitting  the model to
the data obtained from observations.
(iv) Extrapolation of the behaviour to the future
by using the model with the fitted parameters.
These steps are
interconnected and we argue that they should be kept at a similar level of
complexity in order to make their interplay as consistent as
possible. 
In this paper, we investigate whether or not
such consistency is important for obtaining
reliable predictions by exploring several combinations of strategies for steps (i)-(iii) (see summary in Table \ref{Table_1}).

\begin{figure*}
{\includegraphics[clip=true,width=17.5cm]{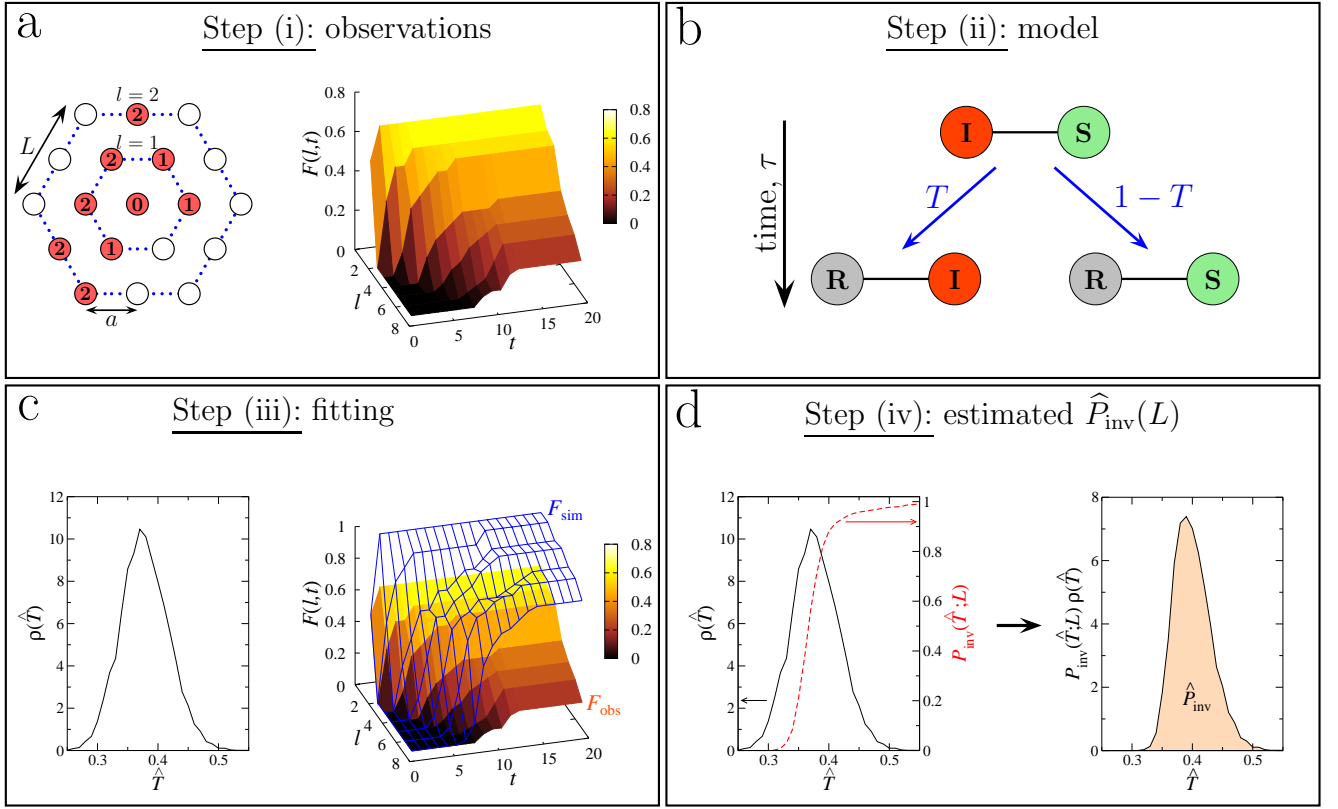}}
\caption{ \label{fig:1} Steps for prediction used in methodology C (Table~\ref{Table_1}). (a-left) Discrete spatio-temporal observations of evolution (spatio-temporal map)
  of   a hypothetical
  epidemic spreading from the central host in a population of hosts
  (circles) arranged on a triangular lattice with lattice spacing $a$
  and $L$ hosts per side of the hexagonal boundaries of the system. The arrangement of nutrient sites in the fungal invasion experiment analysed below is of this type.
  Numbers inside the circles denote the times, $t=0,1, \ldots$, of infection of hosts by
  the observation time $t_{\text{obs}}=2$.  Empty circles correspond
  to healthy (susceptible) hosts by the same time.  The hexagons with
  dotted lines indicate the shells of hosts at a given chemical
  distance, $l$, to the centre of the system.  (a-right) The
  spatio-temporal evolution of the epidemic is described by the
  shell-evolution function $F(l,t)$ giving the relative number of
  hosts in layer $l$ infected by time $t$. For instance, $F(2,2)=3/12$
  for the epidemic shown in the (a-left) panel.  (b) In  step (ii),
  the epidemic is described in terms of an SIR model. An infected
  host, \textcircled{I}, remains infectious during the infectious period
  $\tau$ which, for simplicity, is taken as being constant over the
  whole population and set as a unit of time, $\tau =1$. After the
  infectious period $\tau$, the host is removed, \textcircled{R}.
  During the time $\tau$, \textcircled{I} can transmit the infection
  to a neighbouring susceptible host, \textcircled{S}, with
  probability $T$ (transmissibility).  Alternatively, \textcircled{I}
  can be removed without passing the infection to \textcircled{S} with
  probability $1-T$.  In the discrete time Reed-Frost (RF) dynamics
  used in our approach, the infection is passed instantaneously from
  \textcircled{I} to \textcircled{S} at $t=\tau$.  (c) In step (iii),
  the fitting procedure consists in finding the probability density 
  function (p.d.f.) $\rho(\widehat{T})$ 
  for transmissibilities $\widehat{T}$ (c-left)  such that a RF process with
   $\widehat{T}$  and shell-evolution function
  $F_{\text{sim}}(l,t)$ give a good description of the observed
  shell-evolution function, $F_{\text{sim}}(l,t)$ (c-right).  For
  visualization clarity in c-right, an example of $F_{\text{sim}}(l,t)$ represented by the blue-grid surface does not fit well the observed shell-evolution
  function, $F_{\text{obs}}(l,t)$ (the shaded surface).  The
  p.d.f. $\rho(\widehat{T})$ is obtained by running many RF epidemics
  with random transmissibility and minimising the parameter $d$ that
  quantifies the difference between the observed and the RF
  shell-evolution functions.  (d) Once $\rho(\widehat{T})$ has been
  obtained (continuous curves in c-left and d-left), the probability
  of an invasive epidemic, $\widehat{P}_{\text{inv}}(L)$, can be calculated by
  Eq.~\eqref{eq.1} which involves the conditional probability of
  invasion $P_{\text{inv}}(\widehat{T};L)$ for any given $\widehat{T}$
  (dashed line in d-left).
  The value of $\widehat{P}_{\text{inv}}(L)$
  is represented graphically in d-right by the area under the curve
  (shaded region) for the function  $P_{\text{inv}}(\widehat{T};L)\rho(\widehat{T})$.
}
 \end{figure*}

For concreteness, we illustrate our prediction methods for the particular case of fungal colony invasion in microcosms comprising populations of nutrient sites (agar dots) \cite{Bailey2000}. In these experiments, the central agar dot in ensembles with the
geometry shown in Fig.~\ref{fig:1} was inoculated by
the soil-borne fungal plant pathogen \emph{R. solani} and the spread of the fungal colony is scored in discrete time steps (e.g. daily).
\begin{landscape}
\begin{table}
\begin{flushleft}
\begin{tabular}[t]{c|c|c|c|l|c}
\hline
Methodology & Step (i): data & Step (ii): model & Step (iii): fitting & Parameters & $d^2$ \\ 
\hline
A & \multirow{2}{*}{Mean field (MF), $C(t)$}      & RF & MD  & (ii) $T,\widehat{\tau}_{\text{exp}}$ & \multirow{2}{*}{$d_c^2 = \sum_{t} [c_{\text{sim}}(t)-c_{\text{obs}}(t)]^2$}\\
B &                                  & RF & ABC & (ii) $T,\widehat{\tau}_{\text{exp}}$. (iii) $\epsilon$ & \\
\hline
C & \multirow{3}{*}{Shell, $F(l,t)$} & RF & MD  & (ii) $T,\widehat{\tau}_{\text{exp}}$ & \\
D &                                  & RF & \multirow{2}{*}{ABC} & (ii) $T,\widehat{\tau}_{\text{exp}}$. (iii) $\epsilon$ & \multirow{2}{*}{$d_f^2 = \sum_{l,t} [F_{\text{sim}}(l,t)-F_{\text{obs}}(l,t)]^2$}\\
E &                                  & CT &                      & (ii) $T, \tau_0, k$. (iii) $\epsilon$ \\
\hline
F & Site, $\{t_i\}$                  & CT & DA-MCMC              & (i) $\{t_i\}$. (ii) $T, \tau_0, k$ & \\
\hline
\end{tabular}
\end{flushleft}
\caption{\label{Table_1} {\bf Overview of methods for prediction.} Combination of possible data sets considered in step (i) and methods used for addressing steps (ii) and (iii) in the prediction process. The first column introduces a label for each set of methods, which are ordered according to their overall expected precision.  The second column gives the format of the data obtained from observations in step (i). The smallest precision of the data corresponds to methods A and B in which only the incidence, $C(t)$, is used. Methods C-E use a limited spatio-temporal knowledge of the evolution of the infection given by the  shell-evolution function $F(l,t)$. Method F uses the time of infection of each host, $\{t_i\}$, which is typically unknown from observations but can be inferred in step (iii). In step (ii), RF and CT are abbreviations for the Reed-Frost and continuous-time model, respectively. In step (iii), we have used several methods for fitting: Minimum Distance (MD), Approximate Bayesian Computation (ABC), and data-augmented Markov Chain Monte Carlo (DA-MCMC). Column five lists the parameters involved in each step for prediction of the fungal invasion in the agar-dot experiment. 
Methods based on the RF dynamics are parametrised by the transmissibility, $T$, and a time scale $\widehat{\tau}_\text{exp}$. The CT dynamics used in methods E and F is parametrised by $T$, a characteristic time $\tau_0$, and a shape parameter for the time-dependence of the transmission of infection, $k$. The RF dynamics corresponds to the limit $k\rightarrow\infty$ of the CT dynamics.
The MD method for fitting consists in minimising the parameter $d^2$ (last column) which measures the difference between observations and numerical simulations. The ABC fitting procedure assumes that a simulated invasion fits well the observed epidemic if $d^2<\epsilon$, where $\epsilon$ is a free parameter. 
As shown in the last column, the definition of $d^2$ depends on the descriptors for observations used in step (i). For methods A and B, $d^2$ is defined in terms of the observed and simulated incidences, $c_\text{obs}(t)=C_\text{obs}(t)/N$ and $c_\text{sim}(t)=C_\text{sim}(t)/N$, normalised to the number of hosts in the population, $N$. In methods C-E, $d^2$ is defined in terms of the shell-evolution function for observations and simulations. 
Sections I and II of \emph{SI Appendix} give more details on the definition of models used in step (ii) and fitting methods used in step (iii).}
\end{table}
\end{landscape}

In the following, we give a general description of the steps for prediction of epidemic invasion with particular assumptions suitable for the analysis of the fungal invasion experiment. The main details of all the methodologies are summarised in Table \ref{Table_1}. The methodology C is mainly used for illustration of our concepts.
Details of other explored methodologies are given in \emph{SI Appendix}. The motivation for choosing methodology C is twofold: (i) it keeps all the steps for prediction of the behaviour at a similar level of complexity, as illustrated by analysis of the fungal colony invasion in the agar-dot experiment; (ii) it is a parsimonious methodology that leads to predictions that are, at least, as robust as (or, arguably, even more robust than) those based on more sophisticated approaches (Table \ref{Table_1}).

{\bf Step (i).} The information that can be extracted
from observation of the time evolution of epidemics is usually limited.
In many cases, the only
available information is the incidence, $C(t)$ (the number of infected hosts), at subsequent
observations, and occasionally
the spatial location of infected hosts is also known, e.g. for
epidemics in populations of plants \cite{Otten_Ecology2003_rates,Kleczkowski_JRSocInterf2007,Gibson_PNAS2004,Gibson2006}.
These limitations have a dramatic
influence on subsequent steps in the
prediction process and
it is  crucial to identify which quantities are sufficient for
  prediction of the catastrophic event (i.e. the probability of an invasive epidemic in our case).
As shown in Table~\ref{Table_1}, we consider three 
types of observations. The first consists of discrete temporal observations of $C(t)$ (Methods A and B in Table~\ref{Table_1}). The second possibility (Methods C-E) considers discrete
spatio-temporal observations giving the evolution of infection at discrete times, $t$, in shells at a `chemical distance' $l$ from the initially inoculated host.  As explained in Fig.~\ref{fig:1}(a), such observations can be properly described in terms of a shell-evolution function $F(l,t)$.  As a third possibility (Method F), we use a method for data augmentation in step (iii) that infers the unobserved time of infection for each host, $\{ t_i\}$ \cite{Gibson_PNAS2004,Gibson2006,Kleczkowski_JRSocInterf2007}.

{\bf Step (ii).} We describe the evolution of the epidemic in terms of
a spatial SIR (susceptible-infected-removed) epidemiological model
where the hosts can be either susceptible (S), infected (I) or removed (R)
\cite{grassberger1983,AndersonMay_Book1991,Davis_Nature2008,Otten_Ecology2003_rates}.
This is a prototype model for a wide class of epidemics where disease leads to permanent immunity of hosts after recovery or death. In particular, this paradigm has been shown to be appropriate for description of fungal invasion \cite{Bailey2000,Otten2004,Neri_PLoSCBio2011}. 
In principle, a continuous-time dynamic model is necessary to provide a precise description of epidemic evolution characterised by stochasticity in times of infection and removal/recovery of hosts. Following this idea, it would be natural to use a model with continuous-time (CT) dynamics (described in \emph{SI Appendix}). The drawback of this approach is that it requires knowledge of the  precise times of infection of hosts, $\{t_i\}$, which are typically not available from discrete spatio-temporal observations in step (i).
In order to match an appropriate model with the level of detail of  observations,  we consider the discrete-time dynamics model which
reduces the SIR framework to the so-called Reed-Frost (RF) model
\cite{Daley_BookEpidemicModelling}.
This simplified description is
not expected to capture the dynamical details of the evolution of the
epidemic but reproduces well  its final state~\cite{Ludwig_MBiosc1975,Pellis_MBiosc2008}.
This is a very
important consequence of the fact that, no matter how complicated the
evolution of the epidemic is, the final state of an epidemic with death of infected individuals or
permanent acquired immunity after recovery depends only on
the probability $T$, called transmissibility,
 that the infection has ever been passed between each pair of
  connected hosts (as shown in Fig.~\ref{fig:1}(b)).
Although the transmissibility is expected to exhibit a certain degree of
spatial heterogeneity in real epidemics, we make the minimal
assumption that the trend of the epidemic can be well
  approximated by a RF process with a homogeneous effective
transmissibility $T$.

{\bf Step (iii).} The goal of this step is to estimate the values of the parameters of the model used in step (ii) that give a good description of the
observations. Consider, for definiteness, methodology C in Table~\ref{Table_1}.
Due to
factors such as stochasticity and heterogeneity in transmission, a given
observed spatio-temporal map for infection can occur for different
values of the estimated transmissibility, $\widehat{T}$.
However, some of these values for $\widehat{T}$ are more likely to
produce the observed spatio-temporal pattern than others.
To account for this,
we introduce the probability density function (p.d.f.),
$\rho(\widehat{T})$, which quantifies the probability that the
observed spatio-temporal pattern is reproduced by a certain
value of $\widehat{T}$.
As shown schematically in Fig.~\ref{fig:1}(c),
$\rho(\widehat{T})$ is calculated by generating a large number of stochastic
realizations of the RF epidemic with
transmissibilities $T$ sampled
uniformly from the interval $[0,1]$ and comparing their
shell-evolution functions,
$F_{\text{sim}}(l,t)$ (see caption to Fig.~\ref{fig:1} for definition), with the observed one, $F_{\text{obs}}(l,t)$.
Ideally, the distribution $\rho(\widehat{T})$ would
correspond to the histogram of values for $\widehat{T}$ producing
shell-evolution functions $F_{\text{sim}}(l,t)$ identical to
$F_{\text{obs}}(l,t)$ but obtaining an exact match is computationally very demanding.
Moreover,  reproducing $F_{\text{obs}}(l,t)$ by a RF model is in
general impossible in realistic epidemics for which time is not
discrete.
Therefore, we use a minimum distance (MD) algorithm to calculate $\rho(\widehat{T})$ approximately as the histogram
of values of $\widehat{T}$ minimising the quantity
$d^2_f$ defined in Table~\ref{Table_1} that measures the distance between $F_{\text{sim}}(l,t)$ and $F_{\text{obs}}(l,t)$.
In
this way, the sampled values of  $\widehat{T}$  reproduce
$F_{\text{obs}}(l,t)$ approximately rather than necessarily with distance
$d^2_f=0$. 
This approach is similar to the Approximate Bayesian Computation (ABC) method that determines $\rho(\widehat{T})$ as the histogram of values of $\widehat{T}$ for which $d^2_f \le \epsilon$, where $\epsilon$ is a parameter used in the method \cite{Marjoram_2003:ApproxMCMC}.  Both ABC and MD algorithms give similar results despite the fact that MD does not require the use of an additional parameter $\epsilon$. 
In addition to these approximate methods, we have fitted the spatio-temporal evolution of the CT model proposed in step (ii) for comparison by means of a more standard Bayesian procedure using Markov chain Monte Carlo (MCMC) method with data augmentation (method F in Table~\ref{Table_1}).

{\bf Step (iv).} In the final stage of the prediction process, given $\rho(\widehat{T})$, we
evaluate the probability $\widehat{P}_{\text{inv}}(L)$ that the observed
epidemic will ever invade a system of size $L$.
The epidemic is
defined as being invasive if the final cluster of removed hosts has
reached at least one node on each of the six
 edges of the system.
Otherwise, the epidemic is classified as being
non-invasive.
The conditional probability of invasion $P_{\text{inv}}(\widehat{T};L)$
in a system of size $L$ by an SIR process with \emph{a given
  transmissibility}, $\widehat{T}$,
can be calculated numerically by running many stochastic
realisations of the epidemic and counting the
fraction of invading events.
As shown in Fig.~\ref{fig:1}(d),
$P_{\text{inv}}(\widehat{T};L)$ exhibits a sigmoidal dependence on $\widehat{T}$
which indicates a non-invasive (invasive) regime of epidemics for
relatively small  (large) values of $\widehat{T}$.
Once
$P_{\text{inv}}(\widehat{T};L)$ and $\rho(\widehat{T})$ are known, the
estimated probability of invasion can be  calculated as follows:
\begin{equation}
\label{eq.1}
  \widehat{P}_{\text{inv}} (L)= \int_0^1 P_{\text{inv}} (\widehat{T};L) \rho(\widehat{T})
  \text{d}\widehat{T}~.
\end{equation}
This formula defines the probability that the invasion occurs
given our knowledge about the effective transmissibility encoded by
$\rho(\widehat{T})$ (see a simple graphical
interpretation in terms of the shaded area in  Fig.~\ref{fig:1}(d)).
Importantly,  Eq.~\eqref{eq.1}
gives an extrapolation of the behaviour of the epidemic to
its final state without necessarily providing a
detailed description of the actual evolution leading to such a state.

\section{Application to fungal invasion}
\label{sec:results_fungal}

\begin{figure}

\begin{center}
{\includegraphics[clip=true,width=14cm]{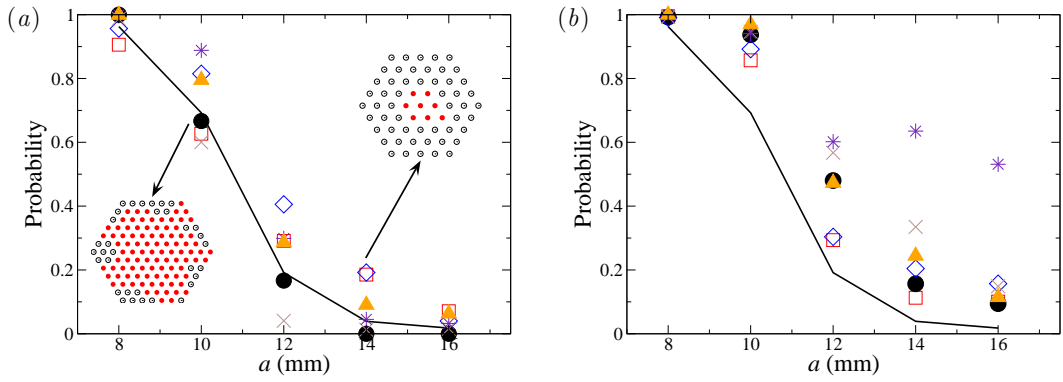}}
\end{center}

\caption{ \label{fig:4} Fungal invasion in the system of agar
    dots placed on a triangular lattice.
(a) The observed
  mean probability of invasion $P_{\text{exp}}$ obtained by counting the
  relative number of
  invasive epidemics after 21~days is shown by the
  continuous  line. The 
 probability of invasion after 21~days was estimated for each replicate by
  observing the initial evolution of colonisation during $t_{\text{obs}}=10$~days. The corresponding mean over replicates with the same value of $a$ is shown with a different symbol type (the same as in Figs.~\ref{fig:rhoT_Average_SD_tobs21}-\ref{fig:Distances_tobs21days}) for each method for prediction. 
  The inserts show the invasive (left) and non-invasive (right)
   state of the epidemic after 21~days for two
  representative replicates with lattice spacings $a=10$~mm and
  $a=14$~mm, as marked by arrows. Solid (open) circles in the inserts represent
  colonised (not colonized) dots. 
(b) Prediction of $\widehat{P}_{\text{inv}}$ with different methodologies for individual
  replicates of the epidemic in a large system of
  size $L=51$ obtained from observations during
  $t_{\text{obs}}=21$~days of the smaller experimental system ($L<8$). The mean of $\widehat{P}_{\text{inv}}$ over replicates for
  each value of $a$ is shown by different symbol types corresponding to different methodologies.
The
  mean probability of invasion $P_{\text{exp}}$ obtained by counting the
  relative number of
  invasive epidemics after 21~days is shown by the
  continuous  line.
}
\end{figure}

In the fungal invasion experiments, the
spatio-temporal maps of infected agar dots were scored daily over 21 days (see two typical patches of
colonisation after $21$~days in Fig.~\ref{fig:4}(a)).
The transmissibility in this experiment corresponds to the probability of
fungal colonisation between two adjacent agar dots and it was
controlled by variable lattice spacing,
$a=8,10,12,14,16,18~\text{mm}$.
Clearly, the experimental setup is restricted both in space and time.
Our aim
is to use these limited observations to estimate the
probability of invasive epidemics in larger systems and for longer times.
The analysis is performed for each individual realisation of the
experiment (6 replicates per value of $a$).

%

In order to make a proper comparison between the experimental observations with the RF model used in methods A-D, 
it is necessary to rescale the time step of
the RF dynamics with dimensionless $\tau=1$ to $\widehat{\tau}_{\text{exp}}$ measured in
days.
The value of $\widehat{\tau}_{\text{exp}}$ is not known and it is treated at the same
level as the transmissibility.
More explicitly, we deal with a bi-variate
probability density function, $\rho_2(\widehat{T},\widehat{\tau}_{\text{exp}})$,
which can be determined for each epidemic with a simple extension of
the methods explained in Sec.~\ref{Sec:Methods} (step (iii)) for obtaining $\rho(\widehat{T})$.
The estimated $\widehat{P}_{\text{inv}}$ is obtained from Eq.~\eqref{eq.1} by defining
$\rho(\widehat{T})$ as the marginal
p.d.f., $\rho(\widehat{T})=\int_0^{\infty}
\rho_2(\widehat{T},\widehat{\tau}_{\text{exp}})
\text{d}\widehat{\tau}_{\text{exp}}$. 
As explained in more detail in Sec. I of \emph{SI Appendix} and summarised in Table~\ref{Table_1}, the continuous-time SIR model used in methods E and F involves three parameters: $T$,$\tau_0$, and $k$. The fitting of the data results in a p.d.f. $\rho_3(T,\tau_0,k)$ from which we obtain $\rho(\widehat{T})=\int_0^{\infty}
\rho_3(\widehat{T},\tau_0,k) \text{d}\tau_0 \text{d} k$. The probability of invasion is then calculated from Eq.~\eqref{eq.1} in the same way as for methods A-D.

\subsection{Uncertainty of the estimated transmissibility}
\label{subsec:uncertainty_T}

\begin{figure}

\begin{center}
{\includegraphics[clip=true,width=6cm,angle=-90]{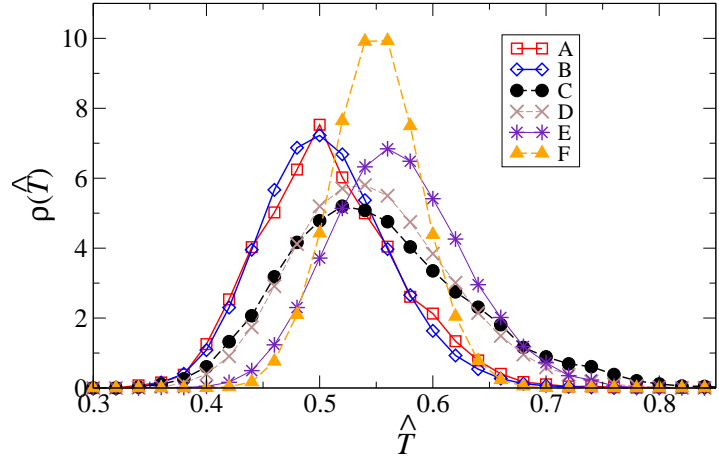}}
\end{center}

\caption{\label{fig:rhoT_a10mm_rep1_tobs21} Estimates of the transmissibility for fungal invasion in the system of agar dots.  The p.d.f.'s $\rho(\widehat{T})$ obtained with different fitting methodologies are plotted for the fungal colony invasion in a population of agar dots with lattice spacing $a=10$~mm. Estimates correspond to observation of the fungal spread during $t_\text{obs}=21$~days. Different symbol types correspond to different fitting methodologies, as marked in the legend. Fig.~S1 in \emph{SI Appendix} shows similar plots for six replicates of the system.}
\end{figure}

The functions $\rho(\widehat{T})$ obtained for the fungal invasion experiments typically exhibit a pronounced peak (see the results for one replicate in Fig.~\ref{fig:rhoT_a10mm_rep1_tobs21} and similar results for more replicates in Fig. S1 of \emph{SI Appendix}). The peaked shape of $\rho(\widehat{T})$ suggests that $\widehat{T}$ can be suitably described in terms of its mean value $\langle \widehat{T} \rangle$ and standard deviation $\sigma_T = (\langle \widehat{T}^2 \rangle - \langle \widehat{T} \rangle^2)^{1/2}$. 
For each methodology, Fig.~\ref{fig:rhoT_Average_SD_tobs21} shows the average over replicates of $\langle \widehat{T} \rangle$ and $\sigma_T$ as a function of the lattice spacing. These estimates correspond to observations of the evolution of infection during $t_\text{obs}=21$~days. All the methods give similar values for  $\langle \widehat{T} \rangle$ which have a clear and expected tendency to decrease with increasing $a$. The uncertainty in $\widehat{T}$, quantified by $\sigma_T$, exhibits greater variations between methodologies but it takes values that are smaller than $\langle \widehat{T} \rangle$ for all the methods and lattice spacings (Fig.~\ref{fig:rhoT_Average_SD_tobs21}(b)). 
This means that $\langle \widehat{T} \rangle$ is a good measure of the typical value of the transmissibility.  
However, the value of $\langle \widehat{T} \rangle$ on its own does not necessarily provide a good approximation for $\widehat{P}_{\text{inv}}$ because the width of $\rho(\widehat{T})$ can bring a significant contribution to the integral in Eq.~\eqref{eq.1}. This is explicitly shown in Sec.~\ref{sec:results_numerical}.

%
%

Comparison of $\langle \widehat{T} \rangle$, $\sigma_T$, and $\rho(\widehat{T})$ for different methods leads to the following conclusions:

\begin{description}
\item{(1)} Given a level of description (step (i)) and a model (step (ii)), the estimates of the transmissibility obtained with ABC and MD methods are, in general, in good agreement (cf. method A with method B, and method C with method D in Figs.~\ref{fig:rhoT_a10mm_rep1_tobs21} and \ref{fig:rhoT_Average_SD_tobs21}).

\item{(2)} Given a level of description (step (i)) and an estimation method (step (iii)), the posteriors obtained using discrete- and continuous-time models are in reasonable agreement (cf. method D with method E in Figs.~\ref{fig:rhoT_a10mm_rep1_tobs21} and \ref{fig:rhoT_Average_SD_tobs21}).
The only difference is a trend for $\rho(\widehat{T})$ corresponding to CT dynamics to have a ``heavy tail'' for large values of $\widehat{T}$ (see the replicate 4 in Fig.~S1, \emph{SI Appendix}). 
Large values of $\widehat{T}$ are correlated with large values of the time scale $\tau_0$ (i.e., slower processes with high $\widehat{T}$) and small values of the shape parameter $k$, that are ruled out by the RF model. This effect becomes more important for larger values of the lattice spacing, as indicated by the large values of $\langle T \rangle$ and $\sigma_T$ corresponding to method E (asterisks in Fig.~\ref{fig:rhoT_Average_SD_tobs21}).

%
%
\begin{figure}

\begin{center}
{\includegraphics[clip=true,width=12cm]{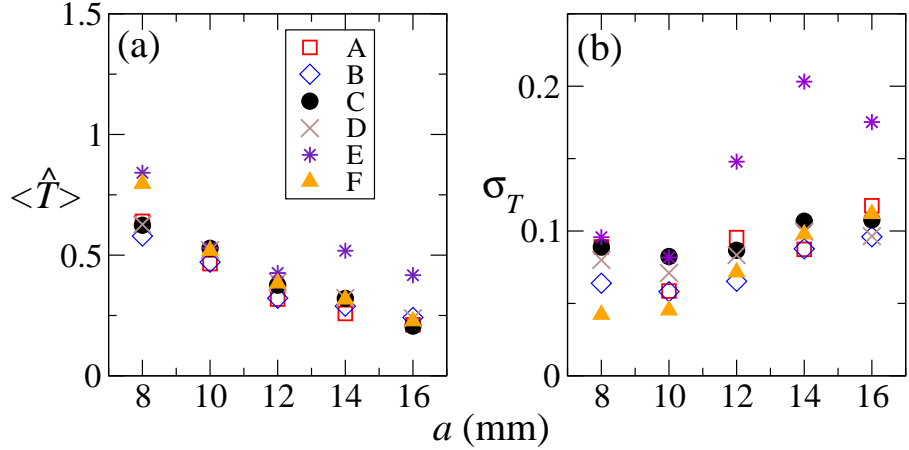}}
\end{center}

\caption{\label{fig:rhoT_Average_SD_tobs21} Statistical characteristics of the estimates of the transmissibility for fungal invasion in the system of agar dots.  Dependence on the lattice spacing of (a) the mean value $\langle \widehat{T} \rangle$, and (b) standard deviation $\sigma_T$ of the transmissibility calculated from the p.d.f. $\rho(\widehat{T})$ corresponding to observations during $t_\text{obs} = 21$~days. For clarity, each symbol gives the average of (a) $\langle \widehat{T} \rangle$ and (b) $\sigma_T$ over 6 replicates of the experiments for each lattice spacing, $a$. As marked in the legend, different symbol types correspond to different methods for addressing the steps (i)-(iii) summarised in Table~\ref{Table_1}.}
\end{figure}

\item{(3)} The estimates from augmented-data MCMC (methodology F) are in general different from those obtained by other methods.
Moreover, the p.d.f.'s $\rho(\widehat{T})$ obtained with MCMC show no systematic trend with respect to the other methods.
With respect to, e.g., the $\rho(\widehat{T})$ obtained with MD, they can be located at slightly higher (replicates 5 and 6 in Fig.~S1) or lower (replicates 2 and 3) values of $\widehat{T}$, or approximately at the same value (replicates 1 and 4). 
Moreover, the variation in the peak position of $\rho(\widehat{T})$ between different replicates is larger than for the other methods. This suggests that the MCMC method is more sensitive to fine details of the evolution of the epidemic. 
A possible explanation is that augmented-data MCMC involves the inference of the unobserved colonisation times and thus is intrinsically individual-based, in contrast to shell-based (or mean-field) methods, which try to match the colonisation times in an approximate manner only.
\end{description}

\subsection{Comparison of fitted models with experimental data}
\label{subsec:comparison_fit_data}

In order to assess the quality of the assumptions used for estimation, we compare the fitted models with the available experimental data. For methods A-D, we compare
the incidence and shell-evolution function obtained numerically (RF dynamics) with values
for $\widehat{T}$ and $\widehat{\tau}_{\text{exp}}$ sampled from
$\rho_2(\widehat{T},\widehat{\tau}_{\text{exp}})$ estimated by means of the spatio-temporal maps at maximum
observation time $t_{\text{obs}}=21~$days
with the actual incidence and shell-evolution function for each epidemic. 
Similarly, the fits from methods E and F are compared with experimental observations by running numerical epidemics with parameters for the CT dynamics sampled from the p.d.f. $\rho_3(\widehat{T},\tau_0,k)$.
We make a quantitative comparison based on squared distances $d_c^2$ and $d_f^2$  (cf. Table \ref{Table_1}) between simulated epidemics and experimental fungal invasions. More explicitly, we define the root mean square (rms) distances,
\begin{align}
\label{eq:Dc}
\Delta c &= \left( \frac{d_c^2}{\Delta t} \right)^{1/2}~, 
\\
 \Delta F &= \left( \frac{d_f^2}{\Delta t \; l_{\max}} \right)^{1/2}~,
\label{eq:DF}
\end{align}
where $\Delta t=21$~days is the time interval used for calculations of $d_c^2$ or $d_f^2$. The quantity $l_{\max}$ is the maximum chemical distance to the centre of the system of agar dots. Its value decreases with the lattice spacing and ranges from $l_{\max}=2$ for $a=16$~mm to $l_{\max}=8$ for $a=8$~mm \cite{Bailey2000}. From the definition of $d_c^2$ given in Table~\ref{Table_1}, it is easy to see that $\Delta c$ gives the typical deviation of the simulated incidence per unit host at a given time, $c_\text{sim}(t)$, from the observed incidence per host at the same time, $c_\text{obs}(t)$. Similarly, $\Delta F$ gives the typical deviation of the simulated shell-evolution function, $F_\text{sim}$, evaluated at any spatio-temporal coordinates $(l,t)$ from the observed value at the same coordinates.

\begin{figure}

\begin{center}
{\includegraphics[clip=true,width=6cm,angle=-90]{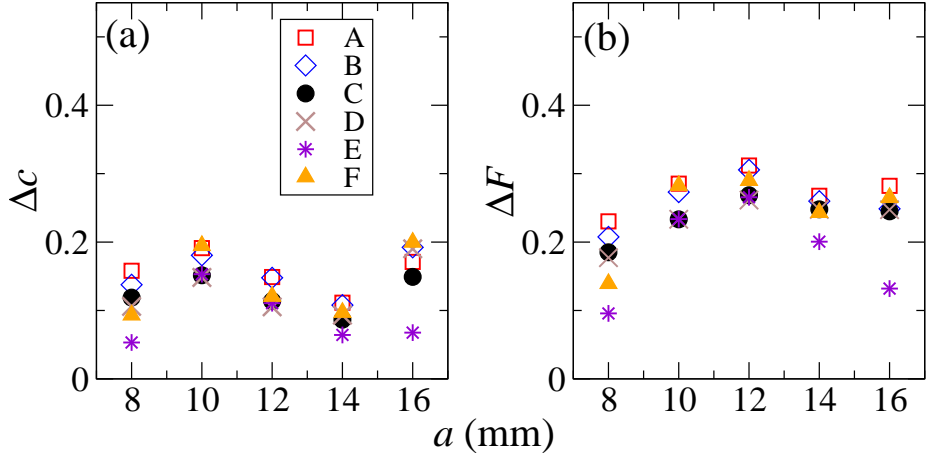}}
\end{center}

\caption{ \label{fig:Distances_tobs21days} Comparison of fitted models with experimental data. Mean rms distances (a) $\Delta c$ and (b) $\Delta F$ between observed and simulated fungal invasions in populations of agar dots. The mean value of rms distances is obtained by averaging over $10^4$ stochastic realisations of simulations and over 6 replicates for each lattice spacing, $a$. Simulations are based on fits to observations over $t_{\text{obs}} = 21$~days. Different symbols correspond to different methodologies summarised in Table~\ref{Table_1}, as marked in the legend.}
\end{figure}

Fig.~\ref{fig:Distances_tobs21days} shows the mean of the rms distances obtained by averaging over stochastic simulations and over replicates with given lattice spacing. The low values of the rms distances ($\Delta c \lesssim 0.2$ and $\Delta F \lesssim 0.3$) indicate that the observed $C(t)$ and $F(l,t)$ are statistically well described by the fitted models. 
For any given method and lattice spacing, we obtained $\Delta c < \Delta F$, which is expected because reproducing the spatio-temporal evolution represented by $F(l,t)$ is more demanding than capturing the temporal evolution of the colonisation given by $c(t)$.
Both $\Delta c$ and $\Delta F$ tend to be larger for $a$ around 10-12~mm which, as shown below, corresponds to cases that are close to the invasion threshold (i.e. where $P_\text{inv}$ decreases from 1 to 0 on increasing $a$, as shown by the continuous line in Fig.~\ref{fig:4}(a)). Variability between replicates of epidemics with given $T$ is larger around the invasion threshold, which is associated with a critical phase transition and characterised by large fluctuations~\cite{grassberger1983,PerezReche_JRSInterface2010,Neri_PLoSCBio2011,Bailey2000,Otten2004,Davis_Nature2008}. As a consequence, the quality of fits is lower in the vicinity of the invasion threshold and this leads to larger values of $\Delta c$ and $\Delta F$.
%
%

Methodology C gives a good balance between performance and number of parameters involved. Methods C-E based on an approximate spatio-temporal description of epidemics given by $F(l,t)$ result in more accurate predictions than methods A (squares in Fig.~\ref{fig:Distances_tobs21days}) and B (diamonds) that neglect spatial features of invasion. 
Moreover, the approximate methods C-E also perform better than even methodology F (triangles in Fig.~\ref{fig:Distances_tobs21days}) despite the fact that the latter aims for a more precise spatio-temporal description. A more qualitative and visual comparison of estimated and observed $C(t)$ reveals similar differences between all the  methodologies (see details in Sec. III of \emph{SI Appendix}).

\subsection{Two applications for prediction methods}
As a first application of the proposed methods, we have studied the predictive power of
the estimates of the probability of invasion and the incidence by
calculating $\rho(\widehat{T})$ from the early stages of the actual
epidemic, i.e. for  $t_{\text{obs}} < 21$~days.  In particular, based on the
estimated $\rho(\widehat{T})$ for $t_{\text{obs}} = 10$~days, we have obtained estimates for the
probability of invasion at time $t=21$~days (squares and dashed line 
in Fig.~\ref{fig:4}(a)) and compared them with the probability $P_{\text{exp}}$
of invasion at $t=21$~days obtained directly from the experimental data. The observed probability of invasion, $P_{\text{exp}}$, is estimated by counting (for each $a$) the fraction of
replicates in which the fungus has reached the six outer edges of the
experimental system by $21$~days.
The mean of $\widehat{P}_{\text{inv}}$ averaged over replicates 
with the same value of $a$ (symbols in Fig.~\ref{fig:4}(a)) gives a reasonable 
estimate for the observed mean of $P_{\text{exp}}$ (solid curve in
Fig.~\ref{fig:4}(a)) after $t=21$~days for most of the methods. Overall, the best predictions for $\widehat{P}_{\text{inv}}$ are obtained with methodology C (solid circles in Fig.~\ref{fig:4}(a)).
As expected, $\widehat{P}_{\text{inv}}$ decreases with increasing $a$ for all methodologies,
illustrating the existence of the threshold for epidemics around
$a\simeq 12~$mm~\cite{Bailey2000}.
Similarly, the experimentally observed incidence and shell-evolution function are statistically well
captured by the numerical extrapolation for their simulated counterparts  up to time
$t=21$~days obtained from observations over times $t_{\text{obs}} < 21$~days. A visual illustration of the agreement between the observed and predicted incidence is given in Sec.~III of \emph{Appendix SI}.
Fig.~\ref{fig:Distances_tobs10days} shows the rms distances between the observed and predicted evolutions from day 11 to day 21. The time interval used to calculate $\Delta c$ and $\Delta F$ from Eqs.~\eqref{eq:Dc} and \eqref{eq:DF} is $\Delta t=11$~days. The relative trends of $\Delta c$ and $\Delta F$ between methods and lattice spacings are similar to those reported in Fig.~\ref{fig:Distances_tobs21days} for the comparison between observations and numerical simulations with $t_\text{obs}$. The main difference is that the values of the rms distances corresponding to predictions of the evolution of colonisation (Fig.~\ref{fig:Distances_tobs10days}) are systematically larger than those obtained by simply comparing observed evolutions with their respective fittings (Fig.~\ref{fig:Distances_tobs21days}). This is in agreement with the intuitive idea that predicting the \emph{a priori} unknown evolution of a system is more challenging than reproducing a fitted evolution.
%
%
\begin{figure}

\begin{center}
{\includegraphics[clip=true,width=6cm,angle=-90]{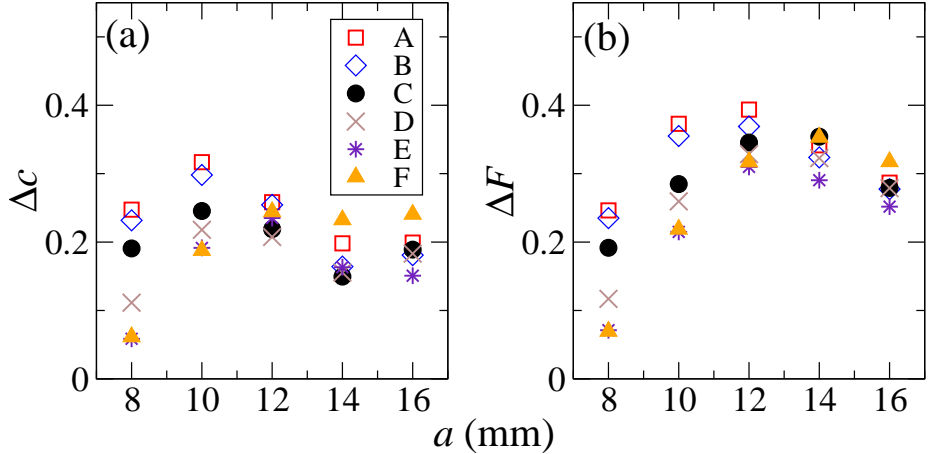}}
\end{center}

\caption{ \label{fig:Distances_tobs10days} Comparison of the predicted and observed evolution of fungal invasion in the system of agar dots. Predictions of the fungal spread in the period of time between days 10 and 21 are made from observation of the initial spread during $t_\text{obs} = 10$~days. The vertical axes show the mean over replicates of the rms  distances (a) $\Delta c$ and (b) $\Delta F$ between observed and predicted fungal evolution during the time interval $11-21$~days. The mean value of rms distances is obtained by averaging over stochastic realisations of simulations and over 6 replicates for each lattice spacing, $a$. Simulations are based on fittings to observations over $t_{\text{obs}} = 10$~days. Different symbols correspond to different methodologies summarised in Table~\ref{Table_1}, as marked in the legend.}
\end{figure}

As a second application of our methodology, we have calculated $\widehat{P}_{\text{inv}}$ at the end of the epidemic as a function of the
lattice spacing in systems of size $L=51$, i.e., larger than the
experimental samples of sizes $L=2,...,8$ which decrease with increasing lattice
spacing (see the two populations for different value of $a$ shown in
Fig.~\ref{fig:4}(a)). Such predictions are based on estimates for the transmissibility obtained from observations up to day $t_\text{obs}$.
As expected, $\widehat{P}_{\text{inv}}$ decreases with increasing $a$,
illustrating the existence of the threshold for epidemics around
$a\simeq 12~$mm~\cite{Bailey2000}.
The results of applying each of the prediction methods are shown in Fig.~\ref{fig:4}(b) for the mean probability averaged over replicates for each value of $a$. 
All the methods except E give similar predictions for $\widehat{P}_{\text{inv}}$. The large values of $\widehat{P}_{\text{inv}}$ predicted by method E are a consequence of the ``heavy tail'' of the p.d.f. $\rho(\widehat{T})$  which gives a significant weigh to the high values of $P_\text{inv}$ for large $\widehat{T}$ in Eq.~\eqref{eq.1}.
The dependence of $\widehat{P}_{\text{inv}}$ on
$a$ differs from the observed probability of invasion, $P_{\text{exp}}$ (continuous curve in
Fig.~\ref{fig:4}(b)).
The difference can be qualitatively understood by 
recalling that $\widehat{P}_{\text{inv}}$ gives an extrapolation both in space and time.
Indeed, $\widehat{P}_{\text{inv}} \geq P_{\text{exp}}$  because some
epidemics that are non-invasive after $21$~days have a certain
probability to invade a system of size $L=51$ for $t>21$~days.
In addition, both infectivity and susceptibility are
expected to be subject to heterogeneity in the agar-dot system due to, e.g.,
inherent variability.  Based on the results presented in Sec.~\ref{sec:results_numerical} for numerical experiments
with heterogeneity in transmission, we expect the estimated $\widehat{P}_{\text{inv}}$
to give an upper bound to the actual probability of invasion. This can
also contribute to the difference between $\widehat{P}_{\text{inv}}$ and $P_{\text{exp}}$ for large
values of $a$.

\section{Numerical experiment}
\label{sec:results_numerical}
The quality of the predictions presented in the previous section is influenced by the quality of the observations in step (i), the suitability of the model chosen in step (ii) for description of the data, and the fitting procedure used in step (iii). In principle, the effect of these factors on predictions could be minimised by optimising the procedures used in each step for prediction. Stochasticity associated with the transmission of infection also influences the ability of making reliable predictions. In contrast to the previous factors, stochasticity is inherent to the nature of the system and its negative effect on predictions cannot be minimised without modifying the system. In this section, we present a sensitivity analysis of our methods by applying them to prediction of invasion for numerically-simulated epidemics where the main factor compromising predictability is the intrinsic stochasticity in transmission of infection. The advantage in this case with respect to more realistic situations is that both the transmissibility, $T$, and the probability of invasion, $P_{\text{inv}}(T;L)$, are known and it is then possible to investigate the performance of the estimates
for $\rho(\widehat{T})$ and $\widehat{P}_{\text{inv}}(L)$ by comparing with the known quantities.

We first consider the simplest situation when the observed
epidemics follow the RF dynamics with homogeneous transmission (i.e. $T$
is the same for all pairs of nearest neighbours in the population).
The idea is to run numerical experiments with known $T$, observe
the evolution of the epidemic over an initial interval of time,
$t_{\text{obs}}$, and then apply the methods described above to
calculate $\widehat{P}_{\text{inv}}(L)$ assuming that $T$ is unknown (as
it occurs for real epidemics).

\begin{figure}
\begin{center}
{\includegraphics[clip=true,width=8.5cm]{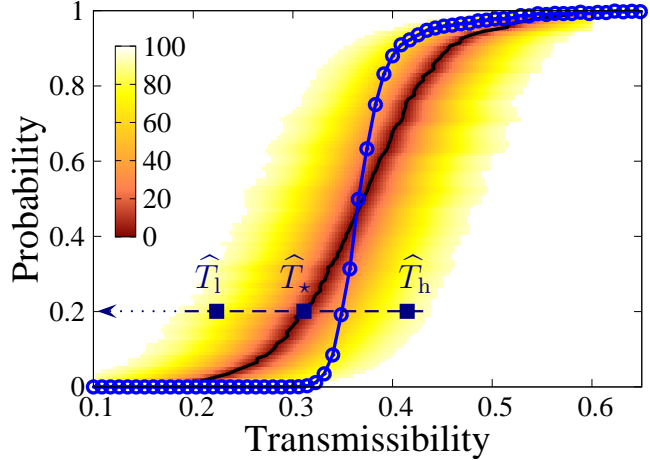}}
\end{center}

\caption{\label{fig:2} {\bf Numerical experiments of SIR epidemics with homogeneous
  transmissibility.} Hosts are placed on the nodes of a triangular
  lattice of size $L=51$ (cf. Fig.~\ref{fig:1}(a)). The
  evolution of epidemics starting from the central host is observed
  over time $t \le t_{\text{obs}} = 7\tau$.  The line marked by circles shows the
dependence of the conditional probability of
invasion, $P_{\text{inv}}(T;L)$, on
transmissibility obtained by the simulations
in the system of size $L=51$.
The shaded region shows the levels of confidence in
percentage of the p.d.f $\rho(\widehat{T})$ around the most
probable transmissibility, $\widehat{T}_{\star}$ (continuous black line)
corresponding to each
value of $\widehat{P}_{\text{inv}}(L)$.
 The horizontal dashed line illustrates the case of estimations giving
$\widehat{P}_{\text{inv}}(L)=0.2$.
If the observed epidemic has a value of the transmissibility
such as $\widehat{T}_{\text{l}}$ that is to the left of the curve for
$P_{\text{inv}}$ (line with circles), the estimated
$\widehat{P}_{\text{inv}}(L)$ overestimates $P_{\text{inv}}$. In
contrast, for values of the transmissibility that are to the right of
the curve for $P_{\text{inv}}$ (e.g. $\widehat{T}_{\text{h}}$), the probability $\widehat{P}_{\text{inv}}(L)$
underestimates $P_{\text{inv}}$.}
\end{figure}

For concreteness, we consider the arrangement shown in Fig.\ref{fig:1}(a) and use methodology C for steps (i)-(iii).
RF epidemics are observed during $t_{\text{obs}}=7\tau$ with the aim of
estimating the probability that they will invade a system of size $L=51$.
Note that over the time interval  $t \le t_{\text{obs}}=7$ the epidemic at most invades a 
hexagon of size $L=15$.
Then, the behaviour of
the epidemic is extrapolated both in space and time. 
We proceed by, first, calculating the p.d.f. $\rho(\widehat{T})$ for
the estimated transmissibilities $\widehat{T}$ compatible with  observation (spatio-temporal map).
Fig.~\ref{fig:1}(c) shows an example of $\rho(\widehat{T})$ obtained from
the analysis of the evolution of an epidemic with $T=0.4$.
In general, the most
probable estimate for the transmissibility, $\widehat{T}_*$, corresponding to the
maximum of $\rho(\widehat{T})$ for a single epidemic, differs from $T$ but not
significantly.
In many cases, $T$ lies within the $68\%$ confidence interval
for $\rho(\widehat{T})$ around its maximum (see a more detailed discussion in
\emph{SI Appendix}). 
The distribution $\rho(\widehat{T})$ allows the probability of invasion $\widehat{P}_{\text{inv}}(L)$ in the system of size $L=51$ to be estimated using Eq.~\eqref{eq.1}.
We have applied this prediction method to many ($\sim 10^4$)
spatio-temporal maps created with  known transmissibility
$T$ spanning the interval $[0,1]$.   For each value of estimated
$\widehat{P}_{\text{inv}}(L)$, the distribution $\rho(\widehat{T})$ is represented by a
horizontal slice of the shaded area in Fig.~\ref{fig:2} (see
e.g. the slice along the dashed blue line corresponding to
$\widehat{P}_{\text{inv}}(L)=0.2$ with darker colour corresponding to higher
probability relative to the maximum of $\rho(\widehat{T})$).  The black ridge in the shaded area corresponds to the
most probable transmissibility, $\widehat{T}_{\star}$, for each value of $\widehat{P}_{\text{inv}}(L)$.
%

To test the quality of the predictions, the estimate of the
probability of invasion $P_{\text{inv}}(\widehat{T};L)$ is compared 
with the probability $P_{\text{inv}}(T;L)$ that would be obtained if
the exact value of $T$ was known \emph{a priori} (see line marked by
circles in Fig.~\ref{fig:2}).  Making such a comparison we can see
that for epidemics with low transmissibility where invasion is
possible but not highly probable, $\widehat{P}_{\text{inv}}(L)$ overestimates
$P_{\text{inv}}(T;L)$ for most of the possible values of $\widehat{T}$
contributing to $\widehat{P}_{\text{inv}}(L)$ (the shaded area including the ridge
region corresponding to the typical values of $\widehat{T}$ is mainly above
the line marked by circles in Fig.~\ref{fig:2}).  This means that
the estimations are biased upwards and the most likely is that 
the actual probability of invasion will be smaller than predicted. In other words, such predictions will typically give a safe bound for the probability of invasion.
Obviously, there is a non-zero probability that the
observed epidemic has a large value of the transmissibility (such as
$\widehat{T}_{\text{h}}$ in Fig.~\ref{fig:2}).  In this case, $\widehat{P}_{\text{inv}}(L)$
would underestimate the actual probability $P_{\text{inv}}$.  For more invasive
epidemics (i.e. epidemics with $P_{\text{inv}} \gtrsim 0.5$), the shaded area
is mainly below the line marked by the circles in
Fig.~\ref{fig:2} meaning that the predicted probability of
invasion $\widehat{P}_{\text{inv}}$ underestimates $P_{\text{inv}}$ for most of the possible
values for $\widehat{T}$, including the most probable, $\widehat{T}_{\star}$.
In these situations however, both $\widehat{P}_{\text{inv}}$ and $P_{\text{inv}}$ are large and
the predictions allow for a reasonable assessment for invasion to be
done. In Sec.~VI of \emph{SI Appendix} we show mathematically that differences between  $\widehat{P}_{\text{inv}}(L)$ and  $P_{\text{inv}}$ evaluated
at the most probable transmissibility $\widehat{T}_{\star}$  are mainly
dictated by the curvature of $P_{\text{inv}}$ around $\widehat{T}_{\star}$ and
is intrinsically linked to the non-zero width of the
p.d.f. $\rho(\widehat{T})$. 
This general result implies that the biases in the probability of invasion at low and high transmissibility are independent of the fitting method (Table \ref{Table_1}) because all methods lead to a p.d.f. $\rho(\widehat{T})$ with non-zero width.

The results presented above correspond to RF epidemics with
homogeneous transmission.  A similar approach has been used to deal
with more realistic epidemics where the transmission of
infection is heterogeneous due to variability in the infectivity and
the susceptibility of hosts. As already mentioned in the previous section, the estimated
$\widehat{P}_{\text{inv}}(L)$ for such epidemics usually gives a bound to the actual probability of
invasion that is even safer than that obtained for cases with
homogeneous transmissibility (see \emph{SI Appendix} for more detail).

\section{Discussion}

%

The methodology introduced here focuses on the prediction of relatively
simple but important features of epidemics.
This is in contrast to much previous
work  dealing with the prediction of quantitative properties of
epidemics such as the detailed spatio-temporal evolution of the
incidence.
The advantage in dealing with simple characteristics of
epidemics is that they can be more easily predicted in terms of
simplified description of the spatio-temporal evolution.
Our results demonstrate that, under quite general assumptions, it is possible
to give reliable prediction of the final
state of an epidemic with permanent immunisation from the early stage of
its evolution.
Such a prediction is possible even for a single
realisation of an epidemic and thus the framework is relevant to
inherently unique real-world epidemics.
In fact, our approach can be
applied for prediction of epidemics in real systems characterised by a
wide range of space and time scales (e.g. crops) based on micro- or meso-cosm
experiments of finite size and over finite time.

The results obtained for experimental fungal invasion by using approximate methods (C-E in Table~\ref{Table_1}) are more robust than those based on supposedly more precise methodology F.
This might be a consequence of the interplay between the high 
fitting precision for MCMC methods involving data augmentation with a poor description of the actual
dynamics given by the continuous-time model fitted to the
observations.
A model capturing dynamical
details at a level consistent with that offered by the fitting
procedure might exhibit more predictive power than that presented here.
This is an illustration of the importance of keeping all the steps
involved in prediction  at a similar
level of complexity in order to give reliable predictions.
For practical applications, the particular method to be used for obtaining the most reliable prediction depends on the problem in  hand. The general rule that seems to emerge from our analysis is that a reasonable method should use as much informations as available from observations, and avoid inferring data that is not directly available unless this is strictly required by the problem. This is the case for methods C-E in our particular study of fungal colony invasion in the population of agar-dots. Methods A and B use less information than available from observations (i.e. they use $C(t)$ instead of $F(l,t)$) while method F infers information that is not available from observations.

The proposed methods assume that epidemics can be approximately
described by an effective transmissibility that is constant over
time and homogeneous in space. However, their  applicability goes beyond epidemics with constant
transmissibility. In particular, we have shown that the method gives
reliable predictions in the presence of spatial heterogeneity in the
transmission of infection. We expect that the methododology can also be
successfully applied to cases where the transmissibility changes over time
but it remains within the bounds for the effective transmissibility
estimated from the early stage.

We have characterised the final state of epidemics by the probability
of invasion. This quantity is suitable for systems with well-defined
boundaries such as, e.g. the population of agar-dots analysed
above.
In cases where hosts are placed on the nodes of more complex
networks, the boundaries of the
system are not necessarily well-defined~\cite{Barrat_08:book} and it is more
convenient to characterise the final state of epidemics in terms of the mean number
of removed (i.e. ever infected) hosts, $N_{\text{R}}$.
Our approach
also applies to such complex networks.
The formula given by Eq.~\eqref{eq.1} provides an estimated size,
$\widehat{N}_{\text{R}}(L)$, if $P_{\text{inv}}(\widehat{T};L)$ is replaced by the
function $N_{\text{R}}(\widehat{T};L)$ giving the size of SIR epidemics with
given transmissibility, $\widehat{T}$.

Our methods have been applied under the assumption that the network of contacts between hosts remains unchanged during the course of epidemics. Such an approximation has been widely used in the past~\cite{Barrat_08:book} and it is reasonable for cases in which the rate of change of the configuration of contacts is much smaller than the removal rate of hosts (i.e. $\sim \tau^{-1}$ in our notation). This condition is clearly satisfied for the fungal colony invasion of the population of agar-dots considered here and also for many other epidemics associated with pathogens spreading in, e.g., networks of plants~\cite{Otten_Ecology2003_rates}, farms~\cite{Keeling_Science2001}, or airports~\cite{Hufnagel_PNAS2004}. This paradigm is also applicable to the spread of many infections in human populations (for instance, measles or SARS that have recovery periods of the order of few days). In contrast, the dynamics of contacts between humans plays an important role for other infectious diseases such as syphilis with a recovery period of $~100$~days~\cite{Volz_ProcRoySocB2007}. A possible strategy to make predictions in this kind of networks would involve inferring the mixing parameter for contacts (as defined, for instance, in \cite{Volz_ProcRoySocB2007}) based on observations (step (i)) in a similar way as we estimated the parameters of the SIR model in step (iii).

Another interesting task would be to extend the ideas presented here to
deal with epidemics with persistence where immunity after recovery is not permanent
(i.e. recovered hosts can be re-infected). In this case, the simplest
model for description of observations  is the SIS model
(susceptible-infected-susceptible)~\cite{Marro_99:book} and a
possible
quantity to be predicted would be the stationary prevalence
of infection (i.e. the density of infected hosts in the stationary state
reached after a transient~\cite{Marro_99:book}).

Due to stochasticity in transmission of infection, it is not possible to determine the parameters of a model describing an epidemic with absolute certainty even if the epidemic is observed during a long time $t_\text{obs}$ before attempting inference. Furthermore, if it were possible to determine the exact value of the parameters, it would still be impossible to make arbitrarily precise predictions of the evolution of the epidemic in the future or predict with absolute certainty if the epidemic is going to be invasive or not (instead, one has to deal with the probability of invasion). The uncertainty in the prediction of the evolution of epidemics grows monotonically with the look-ahead time (see, e.g. the forecast of the incidence in Sec. IV of \emph{SI Appendix}). There exists a prediction horizon beyond which the uncertainty of predictions of the evolution of the epidemic is too large for predictions to be useful. The location of the horizon is epidemic-dependent and also depends on how precise we want our predictions to be. In contrast, for a pure SIR epidemic, there is no prediction horizon for quantities such as $P_\text{inv}$ or $N_\text{R}$ that only depend on the transmissibility. In other words, different replicates of epidemics with given $T$ will follow different evolutions that have a prediction horizon but will lead to the same $P_\text{inv}$ or $N_\text{R}$ \cite{Ludwig_MBiosc1975,grassberger1983,Pellis_MBiosc2008}. More complex nonlinear dynamics for transmission associated with, for instance, a seasonal component in the transmission rate may lead to chaotic behaviour~\cite{Olsen_Science1990}. Predictability of catastrophic events in systems exhibiting chaotic behaviour is a non-trivial question that has been widely studied in the past~\cite{Abarbanel_RMP1993} and still receives considerable attention at present~\cite{Wang_PRL2011}. Even in the absence of stochasticity, the prediction horizon in these systems is intrinsically limited due to the high sensitivity of chaotic processes to the initial conditions and the values of the parameters. In addition, the accuracy of predictions does not necessarily increase monotonically with the observation time, $t_\text{obs}$, before prediction~\cite{Parunak_PredictionHorizon2008}. In such situation, it would be necessary to estimate the value of $t_{\text{obs}}$ leading to the most reliable prediction. Due to all these factors, the methods proposed in this paper may not work when applied for prediction of catastrophic events in nonlinear dynamical systems. However, the ideas presented here together with approaches proposed for prediction in nonlinear dynamical systems may help in devising strategies for prediction in stochastic non-linear systems.

\section*{Acknowledgements}
  The authors acknowledge helpful discussions with G.J. Gibson and
  funding from BBSRC (Grant
  No. BB/E017312/1). CAG acknowledges support of a BBSRC Professorial
  Fellowship.


\begin{thebibliography}{10}

\bibitem{Sornette2000}
Sornette D
\newblock (2000) \emph{Critical Phenomena in Natural Sciences}, Springer Series
  in Synergetics
\newblock (Springer Verlag, Berlin).

\bibitem{Sornette_PNAS2002}
Sornette D
\newblock (2002) Predictability of catastrophic events: Material rupture,
  earthquakes, turbulence, financial crashes, and human birth.
\newblock \emph{Proc Natl Acad Sci USA} 99:2522 -- 2529.

\bibitem{Sornette_StockCrash2003}
Sornette D
\newblock (2003) \emph{Why Stock Markets Crash}
\newblock (Princeton University Press, Princeton).

\bibitem{Medley_Science2001}
Medley GF
\newblock (2001) {EPIDEMIOLOGY: Predicting the Unpredictable}.
\newblock \emph{Science} 294:1663--1664.

\bibitem{Valleron_Science2001_vCJDPrediction}
Valleron AJ, Boelle PY, Will R, Cesbron JY
\newblock (2001) {Estimation of Epidemic Size and Incubation Time Based on Age
  Characteristics of vCJD in the United Kingdom}.
\newblock \emph{Science} 294:1726--1728.

\bibitem{HuillardAignaux_Science2001}
d'Aignaux JNH, Cousens SN, Smith PG
\newblock (2001) {Predictability of the UK Variant Creutzfeldt-Jakob Disease
  Epidemic}.
\newblock \emph{Science} 294:1729--1731.

\bibitem{Morton_JRoySocC2005}
Morton A, Finkenstädt BF
\newblock (2005) Discrete time modelling of disease incidence time series by
  using markov chain monte carlo methods.
\newblock \emph{J. R. Stat. Soc. C} 54:575--594.

\bibitem{Kleczkowski_JRSocInterf2007}
Kleczkowski A, Gilligan C
\newblock (2007) {Parameter estimation and prediction for the course of a
  single epidemic outbreak of a plant disease}.
\newblock \emph{J. Roy. Soc. Interface} 4:865--877.

\bibitem{Keeling_Science2001}
Keeling MJ, {et~al.}
\newblock (2001) {Dynamics of the 2001 UK Foot and Mouth Epidemic: Stochastic
  Dispersal in a Heterogeneous Landscape}.
\newblock \emph{Science} 294:813--817.

\bibitem{Ferguson_Science2001_FootMouthPred}
Ferguson NM, Donnelly CA, Anderson RM
\newblock (2001) {The Foot-and-Mouth Epidemic in Great Britain: Pattern of
  Spread and Impact of Interventions}.
\newblock \emph{Science} 292:1155--1160.

\bibitem{Hufnagel_PNAS2004}
Hufnagel L, Brockmann D, Geisel T
\newblock (2004) {Forecast and control of epidemics in a globalized world}.
\newblock \emph{Proc Natl Acad Sci USA} 101:15124--15129.

\bibitem{Riley_Science2007}
Riley S
\newblock (2007) {Large-Scale Spatial-Transmission Models of Infectious
  Disease}.
\newblock \emph{Science} 316:1298--1301.

\bibitem{PerezReche_JRSInterface2010}
P\'{e}rez-Reche FJ, Taraskin SN, Costa LdF, Neri FM, Gilligan CA
\newblock (2010) {Complexity and anisotropy in host morphology make populations
  less susceptible to epidemic outbreaks}.
\newblock \emph{J. Roy. Soc. Interface} 7:1083--1092.

\bibitem{Truscott_PNAS2003}
Truscott JE, Gilligan CA
\newblock (2003) Response of a deterministic epidemiological system to a
  stochastically varying environment.
\newblock \emph{Proc. Natl. Acad. Sci. USA} 100:9067--9072.

\bibitem{Dye_Science2003_Perspective}
Dye C, Gay N
\newblock (2003) {EPIDEMIOLOGY: Modeling the SARS Epidemic}.
\newblock \emph{Science} 300:1884--1885.

\bibitem{May_Science2004}
May RM
\newblock (2004) {Uses and Abuses of Mathematics in Biology}.
\newblock \emph{Science} 303:790--793.

\bibitem{Pearce_IntJEpidemiol2006_ComplexEpidemics}
Pearce N, Merletti F
\newblock (2006) Complexity, simplicity, and epidemiology.
\newblock \emph{Int. J. Epidemiol.} 35:515--519.

\bibitem{Goldenfel_Science1999_Complexity}
Goldenfeld N, Kadanoff LP
\newblock (1999) Simple lessons from complexity.
\newblock \emph{Science} 284:87 -- 89.

\bibitem{AndersonMay_Book1991}
Anderson RM, May RM
\newblock (1991) \emph{Infectious diseases of humans: dynamics and control}
\newblock (Oxford University Press, Oxford).

\bibitem{Murray_02:book}
Murray JD
\newblock (2002) \emph{Mathematical Biology. I. An Introduction}
\newblock (Springer), 3rd edition.

\bibitem{Davis_Nature2008}
Davis S, Trapman P, Leirs H, Begon M, Heesterbeek J
\newblock (2008) The abundance threshold for plague as a critical percolation
  phenomenon.
\newblock \emph{Nature} 454:634--637.

\bibitem{Otten_Ecology2003_rates}
Otten W, Filipe JAN, Bailey DJ, Gilligan CA
\newblock (2003) Quantification and analysis of transmission rates for
  soilborne epidemics.
\newblock \emph{Ecology} 84:3232 -- 3239.

\bibitem{Bailey2000}
Bailey DJ, Otten W, Gilligan CA
\newblock (2000) Saprotrophic invasion by the soil-borne fungal plant pathogen
  rhizoctonia solani and percolation thresholds.
\newblock \emph{New Phytol.} 146:535 -- 544.

\bibitem{Gibson_PNAS2004}
Gibson GJ, Kleczkowski A, Gilligan CA
\newblock (2004) Bayesian analysis of botanical epidemics using stochastic
  compartmental models.
\newblock \emph{Proc. Natl. Acad. Sci. USA} 101:12120--12124.

\bibitem{Gibson2006}
Gibson GJ, {et~al.}
\newblock (2006) Bayesian estimation for percolation models of disease spread
  in plant populations.
\newblock \emph{Stat Comput} 16:391 -- 402.

\bibitem{grassberger1983}
Grassberger P
\newblock (1983) On the critical behavior of the general epidemic process and
  dynamical percolation.
\newblock \emph{Math. Biosc.} 63:157--172.

\bibitem{Otten2004}
Otten W, Bailey DJ, Gilligan CA
\newblock (2004) Empirical evidence of spatial thresholds to control invasion
  of fungal parasites and saprothophs.
\newblock \emph{New Phytol.} 163:125 -- 132.

\bibitem{Neri_PLoSCBio2011}
Neri F, {et~al.}
\newblock (2011) The effect of heterogeneity on invasion in spatial epidemics:
  from theory to experimental evidence in a model system.
\newblock \emph{PLoS Comput. Biol.} 7:e1002174.

\bibitem{Daley_BookEpidemicModelling}
Daley DJ, Gani J
\newblock (1999) \emph{Epidemic modelling}
\newblock (Cambridge University Press, Cambridge).

\bibitem{Ludwig_MBiosc1975}
Ludwig D
\newblock (1975) Final size distribution for epidemics.
\newblock \emph{Math. Biosc.} 23:33 -- 46.

\bibitem{Pellis_MBiosc2008}
Pellis L, Ferguson NM, Fraser C
\newblock (2008) The relationship between real-time and discrete-generation
  models of epidemic spread.
\newblock \emph{Math. Biosc.} 216:63 -- 70.

\bibitem{Marjoram_2003:ApproxMCMC}
Marjoram P, Molitor J, Plagnol V, Tavar{\'e} S
\newblock (2003) {Markov chain Monte Carlo without likelihoods}.
\newblock \emph{Proc. Natl. Acad. Sci. USA} 100:15324 -- 15328.

\bibitem{Barrat_08:book}
Barrat A, Barth\'{e}lemy M, Vespignani A
\newblock (2008) \emph{Dynamical Processes on Complex Networks}
\newblock (Cambridge University Press, Cambridge).

\bibitem{Volz_ProcRoySocB2007}
Volz E, Meyers LA
\newblock (2007) Susceptible-infected-recovered epidemics in dynamic contact
  networks.
\newblock \emph{Proceedings of the Royal Society B: Biological Sciences}
  274:2925--2934.

\bibitem{Marro_99:book}
Marro J, Dickman R
\newblock (1999) \emph{Nonequilibrium Phase Transitions in Lattice Models}
\newblock (Cambridge University Press, Cambridge).

\bibitem{Olsen_Science1990}
Olsen L, Schaffer W
\newblock (1990) Chaos versus noisy periodicity: alternative hypotheses for
  childhood epidemics.
\newblock \emph{Science} 249:499--504.

\bibitem{Abarbanel_RMP1993}
Abarbanel HDI, Brown R, Sidorowich JJ, Tsimring LS
\newblock (1993) The analysis of observed chaotic data in physical systems.
\newblock \emph{Rev. Mod. Phys.} 65:1331--1392.

\bibitem{Wang_PRL2011}
Wang WX, Yang R, Lai YC, Kovanis V, Grebogi C
\newblock (2011) Predicting catastrophes in nonlinear dynamical systems by
  compressive sensing.
\newblock \emph{Phys. Rev. Lett.} 106:154101.

\bibitem{Parunak_PredictionHorizon2008}
Van Dyke~Parunak H, Belding T, Brueckner S
\newblock (2008) in \emph{Engineering Environment-Mediated Multi-Agent
  Systems}, Lecture Notes in Computer Science, eds{} Weyns D, Brueckner S,
  Demazeau Y
\newblock (Springer Berlin / Heidelberg) Vol.{} 5049, pp 88--102.

\end{thebibliography}

\end{document}



\date{}
\begin{center}
{\bf \huge{Prediction of invasion from the early stage of an epidemic. Supplementary methods and results}}\\
\vskip20pt
{\bf Francisco~J.~P{\'e}rez-Reche$^{1,2}$,
Franco~M.~Neri$^3$,  Sergei~N.~Taraskin$^4$, and
Christopher~A.~Gilligan$^3$}\\
\end{center}
\vskip10pt
\small{$^1$SIMBIOS Centre, University of Abertay Dundee,  Dundee,  UK}\\
\small{$^2$Department of Chemistry, University of Cambridge, Cambridge, UK}\\
\small{$^3$Department of Plant Sciences, University of Cambridge,
Cambridge,  UK}\\
\small{$^4$St. Catharine's College and Department of Chemistry,
University of Cambridge, Cambridge, UK}\\

\tableofcontents
\newpage

\section{Continuous-time epidemiological model for step (ii)}

In this section, we present the three-parameter continuous-time (CT) model used in methods E and F (Table 1, main text) to address step (ii) for prediction. 
In principle, this model is more realistic than the Reed-Frost (RF) model used in methods A-D and allows the existence of possible effects caused by the discrete-time character of the RF description to be explored. 
We use the fungal invasion experiment as a benchmark for the comparison between different models. 
In the CT, the spread (transmission) of the fungal colony between
two neighbouring dots is treated as a time-inhomogeneous Poisson
process \cite{cox1980}. 
The waiting time distribution $f(t)$ for each
transmission event can be modelled by a Weibull
distribution multiplied by the transmissibility $T$: 
\begin{equation}
f(t)=T\frac{t^{k-1}}{\tau_0^k}~e^{-\left(\frac{t}{\tau_0}\right)^k}\, ,
\tag{S.1}
\end{equation}
where $\tau_0$ is the characteristic time scale of the process, and
$k$ is a shape parameter of the distribution. 
Given the cumulative probability function $P(t)=\int_{u=0}^{t}{f(u)\text{d}u}$,
the survival function $S(t)$ (giving the probability that transmission did not occur by time $t$) obeys the following relation:
\begin{equation}
\label{eq:survival}
S(t)=1-P(t)=1-T\left(1-e^{-\left(\frac{t}{\tau_0}\right)^k}\right)\, ,
\tag{S.2}
\end{equation}
The rate of the transmission process,
$\phi(t)$, is a function of the time since colonisation of the donor dot and is given by the expression:
\[
\phi(t)=\frac{f(t)}{S(t)}=-\,\frac{\text{d} \ln\left(S(t)\right)}{\text{d}\,t}~. 
\]
In the limit $k\rightarrow\infty$, the CT model reduces to the RF model, with the same value of $T$ and  infectious period $\tau=\tau_0$. Indeed, for $k\rightarrow\infty$,  the survival probability given by Eq.~\eqref{eq:survival} becomes a step function,
\begin{equation}
S(t)=\begin{cases}
1,& t<\tau_0\\
1-T,& t>\tau_0,
\end{cases}
\nonumber
\end{equation}
which corresponds to a RF model, in which infection can be transmitted from an infected host to a susceptible neighbour with probability $T$ only once the infectious period $\tau_0$ has passed.

\section{Methods for fitting models to data in step (iii)}
The aim of this section is to give a detailed description of the different methods used for fitting models ${\cal M}$ to data ${\cal D}$ described in step (iii) of the methods for prediction proposed in the main text. 
The ideas presented in the main text can be framed within a Bayesian approach which assumes that the parameters $\vtheta$ describing ${\cal M}$ are random variables. The aim of step (iii) is to evaluate $\pi(\vtheta|\D)$, the probability density that ${\cal M}$ with parameters $\vtheta$ describes the data. According to Bayes' rule:
\begin{equation}\label{eq:bayes}
\pi{(\vtheta|\D})\propto~\bP(\D|\vtheta)\pi(\vtheta),
\tag{S.3}
\end{equation}
where $\pi(\vtheta)$ is the prior distribution of the parameters, reflecting our initial belief in their values, and $\bP(\D|\vtheta)$ is the likelihood (the probability of the data given the parameters).

Several challenges arise when using this Bayesian approach in the analysis of epidemic spread.
The first difficulty is associated with inherent limitations in the observations. 
A complete spatio-temporal data set would contain the precise time of colonisation $t_j$ of each host $j$ in the population, i.e. $\D=\{t_j\}$.
Unfortunately, it is often the case that observations do not provide such detailed information. 
In the particular experimental data set considered in the main text, the status of each dot is only recorded at discrete (1-day) time intervals.
Hence, the actual dataset is $\D=\{d_j\}$, where $d_j$ is the day when dot $j$ was first observed as colonised.
We have explored two possible ways to deal with the lack of precise information.
The option used in methods A-E (Table 1, main text) involves identifying descriptors for the evolution of the epidemic that are suitable for the prediction of the catastrophic event.
The lower-dimensional descriptors used in this work are the cumulative incidence, $C(t)$, and the shell-evolution function, $F(l,t)$. 
Another option is to use data augmentation \cite{Gibson1998} that treats the unobserved colonisation times as parameters to be estimated.
This is the procedure followed in method F.

Another source of difficulties is due to the fact that the analytical calculation of the posterior $\pi{(\vtheta|\D})$ is, in general, impossible~\cite{Marjoram_2003:ApproxMCMC}. 
Therefore, it is common to resort to numerical methods to sample from $\pi(\vtheta|\D)$.
In order to do this, we propose a new approximate method (denoted as MD), which is based on the calculation of the minimal distance between two datasets and does not require knowledge of the likelihood $\bP(\D|\vtheta)$. 
In addition, we have used two known methods:
\begin{itemize}
\item a method belonging to the class of Approximate Bayesian Computation (ABC) \cite{Marjoram2003}, which calculates an \emph{approximate} posterior, and which shares several basic features with the MD method.
\item Markov-chain Monte Carlo (MCMC) with data augmentation~\cite{Gibson1998, Gelman2004}, which relies on the exact analytical form of $\bP(\D|\vtheta)$ to sample from the exact posterior $\pi{(\vtheta|\D})$.
\end{itemize}

The ABC method was used with multiple purposes:
\begin{itemize}
\item[(a)] to test the new MD method (involving the \emph{minimisation} of a given distance between observed and simulated data) against an already-known method (ABC, which involves a \emph{cutoff} on the same distance).
\item[(b)] to test different choices in step (ii), by comparing estimations obtained using the RF model with estimations obtained by means of the CT model.
\end{itemize}

The MCMC method was employed to test the new MD method against a widely-used technique that uses both a different level of description (site-level vs. shell or MF level) \emph{and} a different model ${\cal M}$ (CT dynamics instead of RF dynamics). 

\subsection{Minimum distance (MD) method}

Let $\D_\text{obs}$ be the observed data, $\vtheta$ a set of candidate parameters, and $\D_\text{sim}$ a simulated dataset generated using $\vtheta$. Then, if $\D_{\text{obs}}=\D_{\text{sim}}$, the vector of parameters, $\vtheta$, is drawn from $\pi{(\vtheta|\D_\text{obs}})$. In practice, obtaining an exact match between observed and simulated dataset is often computationally unfeasible, and one has to resort to an approximate match. 
To this end, we define a metric $d^2(\D_1,\D_2)$ that measures the distance between two datasets $\D_1$ and $\D_2$. 
The aim of the MD method is to calculate the distribution of parameters $\vtheta$ that minimise $d^2(\D_1,\D_2)$ and gives an approximate posterior. The choice of the metric $d^2\left(\D_{\text{sim}},\,\D_{\text{obs}}\right)$ is problem-specific and, in general, not unique. 
We have tested two different metrics, corresponding to different descriptors of the data (cf. step (i) of the main text):
\begin{enumerate}
\item For the shell-based description, we used $d_f^2=\sum_{l,\,t}\left(F_{\text{sim}}(l,\,t)-F_{\text{obs}}(l,\,t)\right)^2$, where $l$ enumerates the shells, $t$ is the discretised time (observation times in days), and $F$ is the shell-evolution function.
\item Another option is to ignore any spatial information from the data (mean-field (MF) description) and consider only the total fraction of colonised sites at time $t$, $c(t)=C(t)/N$. 
In this case, we chose the distance function $d_c^2=\sum_{t}\left(c_{\text{sim}}(t)-c_{\text{obs}}(t)\right)^2$, where $t$ is again the discretised time.
\end{enumerate}

The algorithm to implement the MD method considers two indexes, $n$, counting the parameter vectors resulting from the minimisation procedure, and $r$, counting the iterations. The values of  $n$  and $r$  are in the range $n,r \in \mathbb{N}$ with maximum values $n_{\text{max}}$ and $R$, respectively, and proceeds as follows:

\begin{itemize}
\item[MD.1] Set $n=0$
\item[MD.2] Set $r=0$
\item[MD.3] Chose a value $\vtheta_n^{(r)}$ for the parameter vector sampled from the prior $\pi(\vtheta)$. 
\item[MD.4] Generate a data set $\D_{\text{sim}}^{(r)}$ from the model ${\cal M}$ with parameters $\vtheta_n^{(r)}$.
\item[MD.5] Calculate $d^2(\D_{\text{obs}},\,\D_{\text{sim}}^{(r)})$ and
\begin{itemize}
\item If $r<R$, set $r=r+1$ and return to MD.3 or
\item If $r=R$, go to MD.6
\end{itemize}
\item[MD.6] Among all the parameters $\{\vtheta_n^{(r)};\, r=0,1,\dots,R\}$, chose the set of parameters $\vtheta_{n}\in \{\vtheta_n^{(r)}\} $ giving the closest simulated data, $\D_{\text{sim}}$ to observations, i.e. $d^2(\D_{\text{obs}},\,\D_{\text{sim}})=\min_{r=0,\dots, R}\{d^2(\D_{\text{obs}},\,\D_{\text{sim}}^{(r)})\}$.
\item[MD.7] Set $n=n+1$ and return to MD.2 until $n=n_{\text{max}}$. 
\end{itemize}

The result of this algorithm is a set of parameter vectors $\{\vtheta_{n};\, n=0,1,\dots,n_{\text{max}}\}$ that give simulated data with minimum distance to observations.
The normalised histogram for the obtained parameter vectors $\{\vtheta_{n}\}$ defines a p.d.f. $\rho(\vtheta)$ that approximates the posterior $\pi{(\vtheta|\D})$.  

For all the results presented in the paper and obtained with the MD algorithm, we set $R=5000$.
For some fungal epidemics, we have checked that larger values of $R$ do not lead to smaller values of  $d^2(\D_{\text{obs}},\,\D_{\text{sim}})$ in step [MD.6].

\subsection{Approximate Bayesian Computation (ABC)}

In common with the MD method, the ABC approximate Bayesian method we have used also relies on the definition of a metric, $d^2$. 
If the distance between observed and simulated datasets is less than a given tolerance parameter,  $\epsilon$, i.e. $d^2(\D_{obs},\,\D_{sim})\leq\epsilon$, then $\btheta$ is drawn from the \textit{approximate} posterior $\pi\left(\btheta|d^2(\D_{obs},\,\D_{sim})\leq\epsilon\right)$. The accuracy of the approximation increases as $\epsilon\rightarrow~0$.

For our estimations, we used a Markov-chain Monte Carlo algorithm that can be summarised as follows:
\setlength{\leftmargini}{3.6em}
\begin{itemize}
		\item[ABC.1] Set $n=0$ and choose the initial value $\vtheta_0$ of the parameter vector.
		\item[ABC.2] Generate a candidate vector, $\vtheta'$, from a proposal distribution $q(\vtheta'|\vtheta_n)$.
		\item[ABC.3] Generate a data set $\D_{\text{sim}}$ from the model ${\cal M}$ with parameters $\vtheta'$.
		\item[ABC.4] Calculate $d^2(\D_{\text{obs}},\,\D_{\text{sim}})$ and
		\begin{itemize}
				\item if $d^2(\D_{\text{obs}},\,\D_{\text{sim}})\leq\epsilon$, go to ABC.5;
				\item if $d^2(\D_{\text{obs}},\,\D_{\text{sim}})>\epsilon$, set $\vtheta_{n+1}=\vtheta_n$ and go to ABC.7.
		\end{itemize}
		\item[ABC.5] Calculate the probability of acceptance:
		\begin{equation}
				p_\text{acc}=\text{min}\left(1,\frac{\pi({\vtheta'})q(\vtheta_n|\vtheta')}{\pi(\vtheta_n)q(\vtheta'|\vtheta_n)}\right),
                                \nonumber 
		\end{equation}
		\item[ABC.6] Set $\vtheta_{n+1}=\vtheta'$ with probability $p_\text{acc}$, or $\vtheta_{n+1}=\vtheta_n$ with probability $1-p_\text{acc}$.
		\item[ABC.7] Set $n=n+1$ and return to ABC.2 until the chain has converged and the required number of samples has been collected.
\end{itemize}

In our estimations, 
either a uniform  distribution with the same support as the prior (see below) or normal distribution, ${\cal N}(0,\,\sigma^2)$ were used for
the proposal distribution $q(.)$ (cf. steps ABC.2 and ABC.5). 
The criterion for the choice between these two distributions was to minimize the average number of simulations needed to generate a dataset with $d^2(\D_{\text{obs}},\,\D_{\text{sim}})\leq\epsilon$ in step [ABC.4].
The value of $\sigma$ for the normal distribution was chosen according to the same criterion, and was typically a fraction between $0.05$ and $0.1$ of the support of the prior distribution.

Every chain was run for $5\times 10^4$ steps, discarding an initial burn-in period of $5\times 10^3$ steps. Since very small values of the tolerance $\epsilon$ imply very low acceptance rates, the final choice of $\epsilon$ was the result of a tradeoff between the accuracy of the approximation and CPU time available. The values of $\epsilon$ that were finally used in our analyses are shown in Table~S\ref{tab:epsilon}.

\begin{table}
	\centering
		\begin{tabular}{| c | c | c | c |}
		\hline
			Rep.	& D 			& B 		& E \\ \hline
			$1$  	& $0.03$  & $1.5$ & $0.5$  \\ 
			$2$   & $0.03$  & $1.5$ & $0.7$  \\ 
			$3$   & $0.03$  & $1.5$ & $1.5$  \\ 
			$4$   & $0.03$  & $1.5$ & $0.5$  \\ 
			$5$   & $0.03$  & $1.5$ & $1$    \\ 
			$6$   & $0.03$  & $1.5$ & $1$    \\ \hline
		\end{tabular}
		\caption{Values of $\epsilon$ used for methods B, D, and E (Table 1, main text) using the ABC inference method in step (iii) for prediction.}
		\label{tab:epsilon}
\end{table}

For both MD and ABC approaches, we assumed independent priors for all the parameters: $\pi(\vtheta)=\pi(T)\pi(\tau_{\text{exp}})$ (RF model) and $\pi(\vtheta)=\pi(T)\pi(\tau_0)\pi(k)$ (CT model). For the RF model, all the priors were uniform: $\pi(T)=U(0,\,1)$ and $\pi(\tau_\text{exp})=U(\tau_\text{min},\,\tau_\text{max})$, where $\tau_\text{min}=1\text{d}$ and $\tau_\text{max}$ was changed between treatments (increasing with the lattice spacing, from $\tau_\text{max}=6\text{d}$ for the lattice with $a=8$mm spacing to $\tau_\text{max}=12\text{d}$ for the lattice with $a=16$mm spacing). For the CT model, we used two sets of priors: (i) noninformative uniform priors for all the parameters ($\pi(T)=U(0,\,1)$, $\pi(\tau_0)=U(0,\,20)$, $\pi(k)=U(0,\,20)$) and (ii) noninformative prior for the transmissibility ($\pi(T)=U(0,\,1)$) and exponential priors for the other two parameters ($\pi(\tau_0)=\text{Exp}(1)$, $\pi(k)=\text{Exp}(1)$ as for the augmented-data MCMC estimations). Only results for noninformative priors from set (i)  are presented below, and compared with results from the RF model. The different choice for the priors from set (ii) has an effect on the posterior distributions, but does not affect significantly the predicted incidence curves (see Section~3) and, for the sake of brevity, results from set (ii) are not presented.

The ABC and MD methods share several common features. Both of them rely on the simulation of epidemics using the current parameters, and on the calculation of a distance between simulated epidemics and the observed data. 
However, the ABC method considers any distance below the cutoff $\epsilon$, while the MD method seeks to minimise the distance over a given number of simulations $R$. The iterative procedure used in MD method allows the use of an additional parameter analogous to $\epsilon$ in ABC method to be avoided. In both cases, the exact posterior is recovered in the proper limit ($\epsilon\rightarrow~0$ and $R\rightarrow\infty$, respectively).

\subsection{Markov Chain Monte-Carlo (MCMC) method with data augmentation}
In order to implement the data-augmented MCMC method, it is necessary to calculate the explicit form of the likelihood $\bP(\D|\vtheta)$ (with $\vtheta=(T,\,\tau_0,\,k)$).
We sketch here the main steps of the calculation.
Let $\I$ be the set of the dots that are colonised before the end of the experiment (at time $\tend=21$~days), and $\U$ be the set of those that are still uncolonised at $t=\tend$. 
Assume first that we know the times of colonisation $t_j$ of each dot $j\in\I$. 
The data then consist of the vector $\mathbf{t}$ of colonisation times, plus the set of uncolonised dots, i.e. $\D=(\mathbf{t},\,\U)$. 
The nearest neighbours of dot $j$ form the set $\N_j$, and the potential donors of $j$ form the subset $\S_j \subseteq \N_j$.
If $j\in\U$, then $\S_j$ contains the colonised neighbouring dots, i.e. $\S_j=\{i:\,i\in\N_j\cap\I\}$.
If in contrast $j\in\I$, $\S_j$ contains the neighbouring dots that are colonised before $j$, i.e. $\S_j=\{i:\,i\in\N_j\cap\I,\,t_i<t_j,\}$.

Given these definitions, the likelihood function can be written as the product of the contributions from individual dots $j$:
\begin{equation}\label{eq:likelihood_all}
\bP(\D|\vtheta)=\prod_{j\in\I}{f^\I_j(t_j)}\prod_{j\in\U}{\P^{\,\U}_j(\tend)},
\tag{S.4}
\end{equation}
where $f^\I_j(t_j)$, is the p.d.f. for the colonisation times $t_j$, and $\P^{\,\U}_j(\tend)$ is the probability for dot $j$ to be uncolonised by the end of the experiment.
These contributions can be calculated explicitly as follows.
The probability that a dot $j\in\I$ has not been colonised by a given neighbour $i\in\S_j$ by time $t_j$ is given by the survival function $S(t_j-t_i)$ (see Eq.~(\ref{eq:survival})). 
Hence, the probability $\P^\I_j(t_j)$ that dot $j$ is still uncolonised at time $t_j$ is given by the product over all $i\in\S_j$:
\begin{equation}\label{eq:survival_I}
\P^\I_j(t_j)=\prod\limits_{i\in~S_j}{S(t_j-t_i)}=\prod\limits_{i\in~S_j}{\text{exp}\left({-\int_{0}^{t_j-t_i}\phi(t)\,\text{d}t}\right)}
=\text{exp}\left(-\sum\limits_{i\in~S_j}{\int_{0}^{t_j-t_i}\phi(t)\,
\text{d}t}\right),
\tag{S.5}
\end{equation}
where we used the relation $S(t)=\text{exp}(-\int_0^t\phi(u)\text{d}u)$.
The p.d.f. for $t_j$ is then given by:
\begin{equation}\label{eq:likelihood_I}
f^\I_j(t_j)=-\frac{d\P^\I_j(t_j)}{dt_j}=\left(\sum\limits_{i\in~S_j}{\phi(t_j-t_i)}\right)
\text{exp}\left(-\sum\limits_{i\in~S_j}{\int_{0}^{t_j-t_i}\phi(t)\,\text{d}t}\right),
\tag{S.6}
\end{equation}
Likewise, a dot $j\in\U$ is still uncolonised by time $\tend$ when transmission did not occur from any of its neighbours $i\in\S_j$, yielding the probability:
\begin{equation}\label{eq:likelihood_U}
P^{\,\U}_j(\tend)= \begin{cases} 
\prod\limits_{i\in~\S_j}{S(\tend-t_i)}~, &~~~~\text{if}~~~~ \S_j \ne \emptyset~, \\
1~, &~~~~\text{if}~~~~ \S_j = \emptyset~.
\end{cases} 
\tag{S.7}
\end{equation}

In general, we are interested in obtaining the marginal distribution of a single parameter (in particular, $T$) from Eq.~(\ref{eq:bayes}), and thus all other parameters entering the expression for $\bP(\D|\vtheta)$ have to be integrated out.
This is, in general, unfeasible analytically.
In the fungal invasion experiment, the calculation is further complicated by \emph{censoring}, i.e., by the fact that experimental observations are made at discrete (1-day) time intervals.
As a consequence, if $d_j$ is the day when dot $j$ was first recorded as colonised, then the colonisation time $t_j$ is constrained to lie in the interval $t_j\in(d_j-1,\,d_j)$, but its exact value is unknown.  
The actual dataset is hence $\D=(\mathbf{d},\,\U)$, and the likelihood has to be calculated from Eqs.~(\ref{eq:likelihood_all}-\ref{eq:likelihood_U}) by integrating out the unobserved colonisation times,
\[
\bP(\mathbf{d},\,\U|\vtheta)=\int_{\mathbf{T}{(\mathbf{d})}} \bP(\mathbf{t},\,\U|\vtheta)\text{d}\mathbf{t}~,
\] 
where the integral is performed over the (high-dimensional and, in general, very complex) space $\mathbf{T}(\mathbf{d})$ compatible with the observed data.

Since the high-dimensional integrals introduced above are analytically intractable, numerical techniques are commonly used to sample values from the posterior distribution $\pi{(\vtheta|\D})$. 
The MCMC method consists in implementing a Markov chain for $\vtheta$ that has $\pi{(\vtheta|\D})$ as stationary distribution. A large literature exists on this subject (see e.g. Ref.~\cite{Gelman2004}), to which the reader is referred for details. 
In our case, a Metropolis-Hastings algorithm has been used to build the Markov chain. 

The algorithm can be summarised as follows:
\setlength{\leftmargini}{4.6em}
\begin{itemize}
		\item[MCMC.1] Set $n=0$ and choose the initial value $\vtheta_0$ of the parameter vector.
		\item[MCMC.2] Generate a candidate  value of the  vector $\vtheta'$ from a proposal distribution $q(\vtheta'|\vtheta_n)$.
		\item[MCMC.3] Calculate the probability of acceptance:
		\begin{equation}
				p_\text{acc}=\text{min}\left(1,\,\frac{\pi{(\vtheta'|\D})}{\pi{(\vtheta_n|\D})}\right)=\text{min}\left(1,\frac{\bP(\D|\vtheta')	\pi{(\vtheta'})q(\vtheta_n|\vtheta')}{\bP(\D|\vtheta_n)\pi(\vtheta_n)q(\vtheta'|\vtheta_n)}\right),
                                \nonumber 
		\end{equation}
		\item[MCMC.4] Set $\vtheta_{n+1}=\vtheta'$ with probability $p_\text{acc}$, or $\vtheta_{n+1}=\vtheta_n$ with probability  $1-p_\text{acc}$.
		\item[MCMC.5] Set $n=n+1$ and return to MCMC.2 until the chain has converged and the required number of samples has been collected.
\end{itemize}
In order to deal with the unobserved colonisation times $\mathbf{t}$, we used data augmentation \cite{Gibson1998}. Their values were treated as parameters to estimate, i.e. the original parameter vector $\vtheta$ was  expanded (augmented) to $(\vtheta,\,\mathbf{t})$.

New colonisation times were then proposed and accepted/rejected within the same Metropolis-Hastings algorithm. 
Additional care had to be taken in this case, since at each step a pathway of transmission must exist between the dot inoculated at time $t=0$ and all the other colonised dots in the system (see the discussion in \cite{Gibson1998} and \cite{Gibson2006}).

Independent priors were used for all the  parameters, so that $\pi(\vtheta)=\pi(T)\pi(\tau_0)\pi(k)$.
A noninformative uniform prior was used for the transmissibility, i.e. $\pi(T)=U(0,\,1)$.
For the other two parameters, exponentially distributed priors were used, i.e. $\pi(\tau_0)=\text{Exp}(1)$, $\pi(k)=\text{Exp}(1)$.
Such choice was made in order to exploit our prior knowledge, i.e., respectively, that the typical time scale of the fungal spread in our system is of the order of days, and that the waiting-time distribution is not step-like (i.e., $k$ is not too large; note that such assumption is opposite to that of the RF model).
We checked that the posterior distribution was not too sensitive to changes up to a factor 2 in the parameters of the exponential priors.
A uniform prior was used for each augmented colonisation time.

Every chain was run for $10^5$ MCMC steps discarding an initial burn-in period of $10^3$ steps.
We checked that the final posterior distribution was robust with respect to the choice of the inital point $\vtheta_0$. 

\subsection{Distribution for the estimated transmissibility, $\rho(\hat{T})$}

This section complements the results presented in the main text (Figs. 2 and 3) for the p.d.f. $\rho(\hat{T})$ obtained with different methods for parameter estimation in the fungal invasion experiment. 
Within the Bayesian framework, the analogous of the distribution $\rho(\hat{T})$ introduced in the main text for the MD method is the marginal p.d.f of the posterior $\pi{(\vtheta|\D})$ integrated over the variable $\tau_\text{exp}$ for the RF model and over the variables $\tau_0$ and $k$ for the CT model.

\begin{figure}
\includegraphics[width=\textwidth,angle=0]{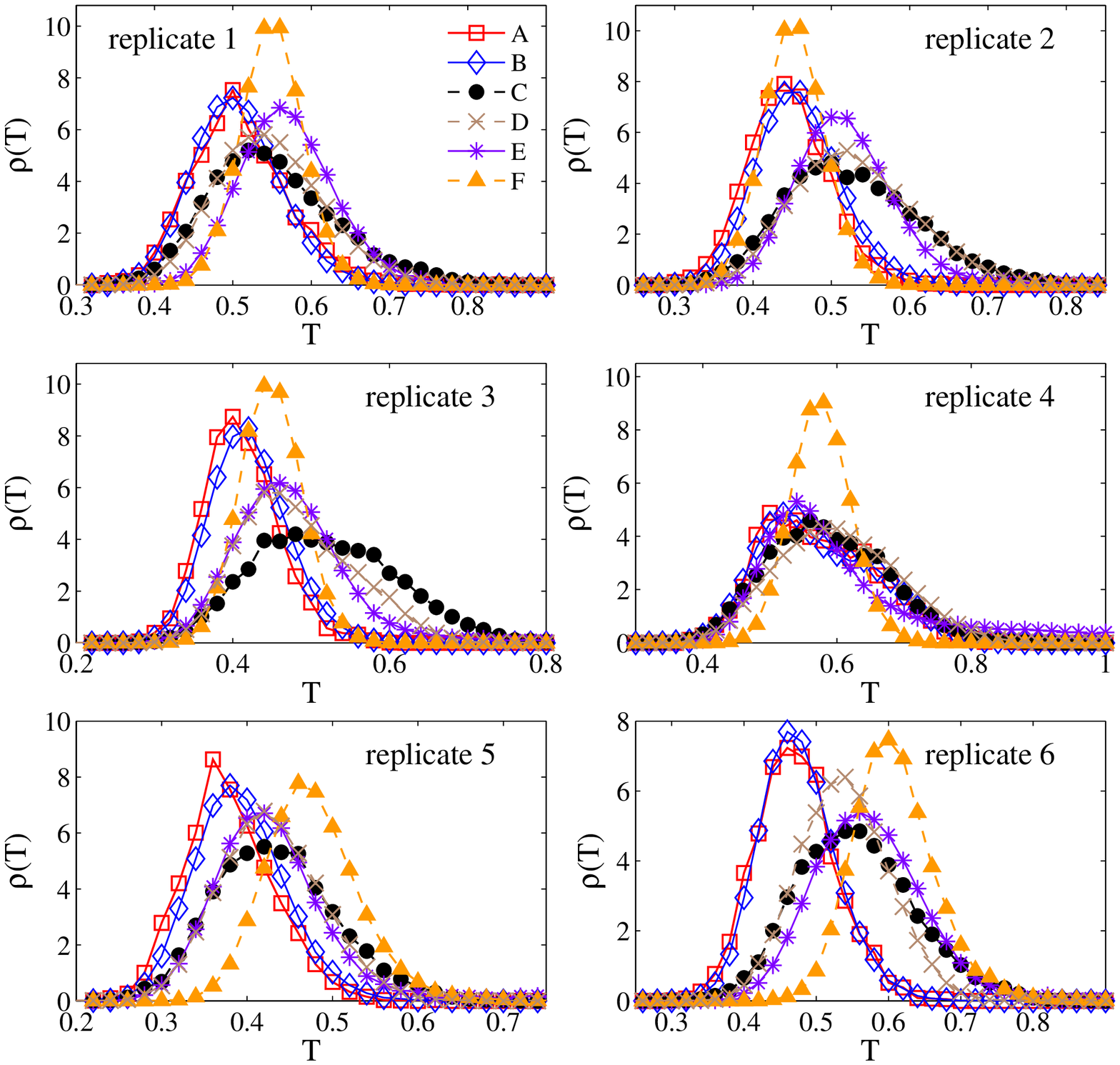}
\caption{\label{fig:rho_10mm} Comparison of marginal posterior distributions $\rho(\hat{T})$ obtained with methods listed in Table 1 of the main text. Different symbols and lines correspond to different methods, as indicated by the legend. All the replicates correspond to invasion in populations with lattice spacing $a=10$~mm.}
\end{figure}

Fig.~S\ref{fig:rho_10mm}  shows the comparison of the probability density functions $\rho(\hat{T})$ obtained for the agar dot experiment by all the methods summarised in Table 1 of the main text.
The posteriors are shown for experiments in populations with lattice spacing $a= 10$mm.
This particular set of results was chosen because it summarised well all the main effects of the methodologies used.
All the estimations correspond to observations over the complete duration of the experiment, i.e. $t \leq t_{\text{obs}}=21$~days.

\section{Additional comparisons of fitted models with experimental data}
\label{sec:comparison_model_experiment}

In the main text, we presented a test of the goodness of fit of models to data based on the mean squared distances $d_c^2$ and $d_f^2$. The purpose of this section is to give a more visual comparison between fitted models and observations based on the cumulative incidence.

For every model used, we obtain the statistics for the incidence corresponding to fitted models by sampling the parameters $\vtheta$ from the joint (exact or approximate) posterior, $\pi{(\vtheta|\D})$. 
This procedure gives a p.d.f. $\rho(C|t)$ for the incidence $C$ at any given time $t$. 
The dispersion of $\rho(C|t)$ is associated both with the stochasticity in the simulated model for each value of the parameters, $\vtheta$,  and the dispersion for the values of these parameters given by $\pi{(\vtheta|\D})$. 
Strictly speaking, the comparison of the experimental incidence with that obtained by methods A-D based on the RF model makes sense only when the stochastic nature of the process is taken into account.
Indeed, the fine details of the two types of processes are different: while the experimental curve $C(t)$ corresponds to a discrete sampling  of a \emph{continuous-time process} (discrete set of observations), $\rho(C|t)$ gives the statistics of effective \emph{discrete-time processes} with random values of transmissibility corresponding to $\vtheta$ being drawn from $\pi{(\vtheta|\D})$. 

\begin{figure}
\includegraphics[width=\textwidth,angle=0]{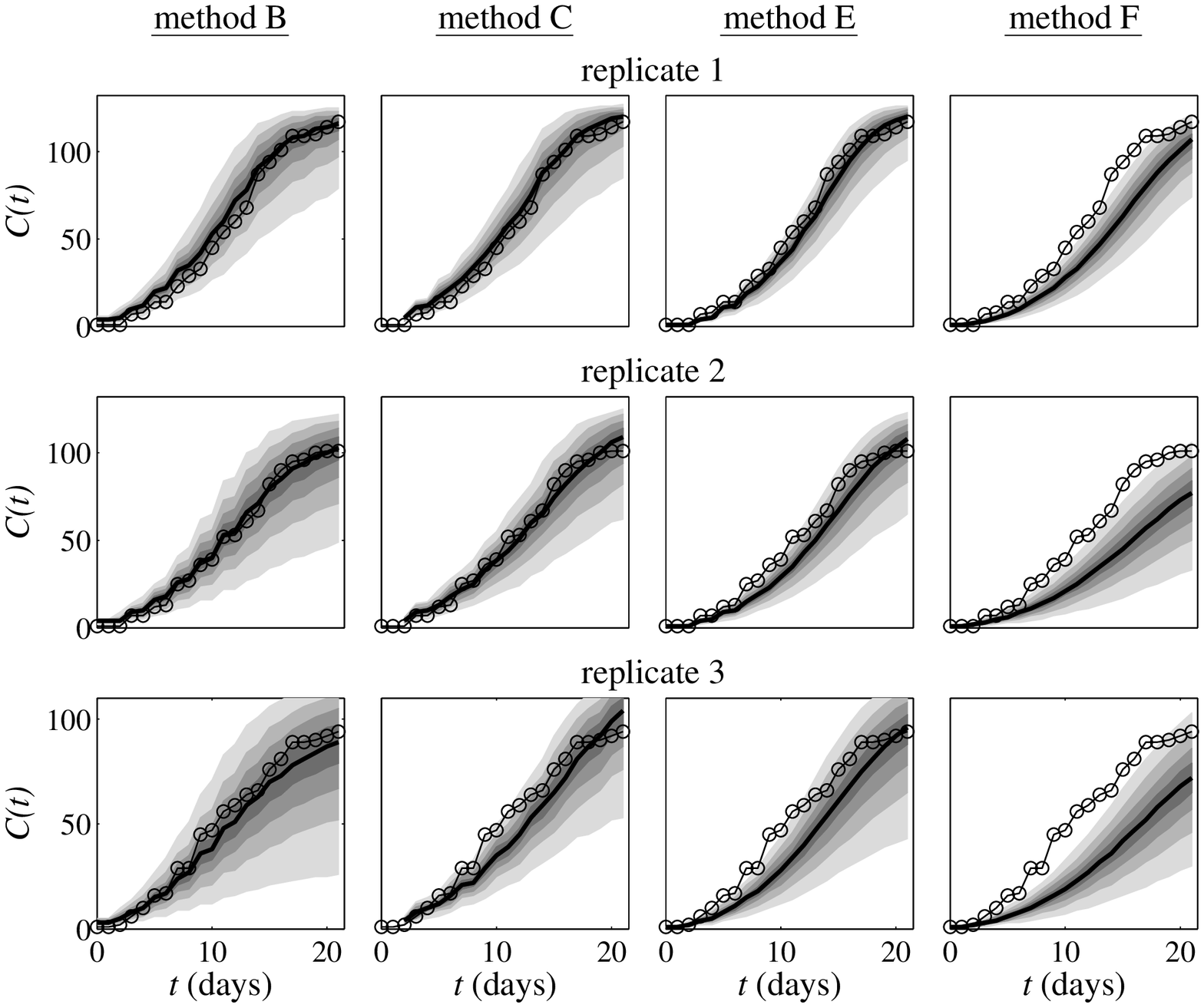}
\caption{\label{fig:DPC_10mm_1}
Incidence (line with circles) for fungal invasion of a set of agar dots arranged on a triangular lattice with spacing $a=10$~mm \cite{Bailey2000}. 
Replicates 1-3 are shown.
For each replicate, four panels are shown, with the p.d.f. $\rho(C|t)$ for the incidence $C$ at any time $t$ obtained by means of methods B, C, E, and F (as described in Table~1 of the main text).
In all the panels, the ridge (bold solid line) corresponds to the median of $\rho(C|t)$.
The grey-scale shaded areas are the $20\%$ (darker), $40\%$, $60\%$, and $80\%$ (lighter) percentiles around the median.}
\end{figure}

\begin{figure}
\includegraphics[width=\textwidth,angle=0]{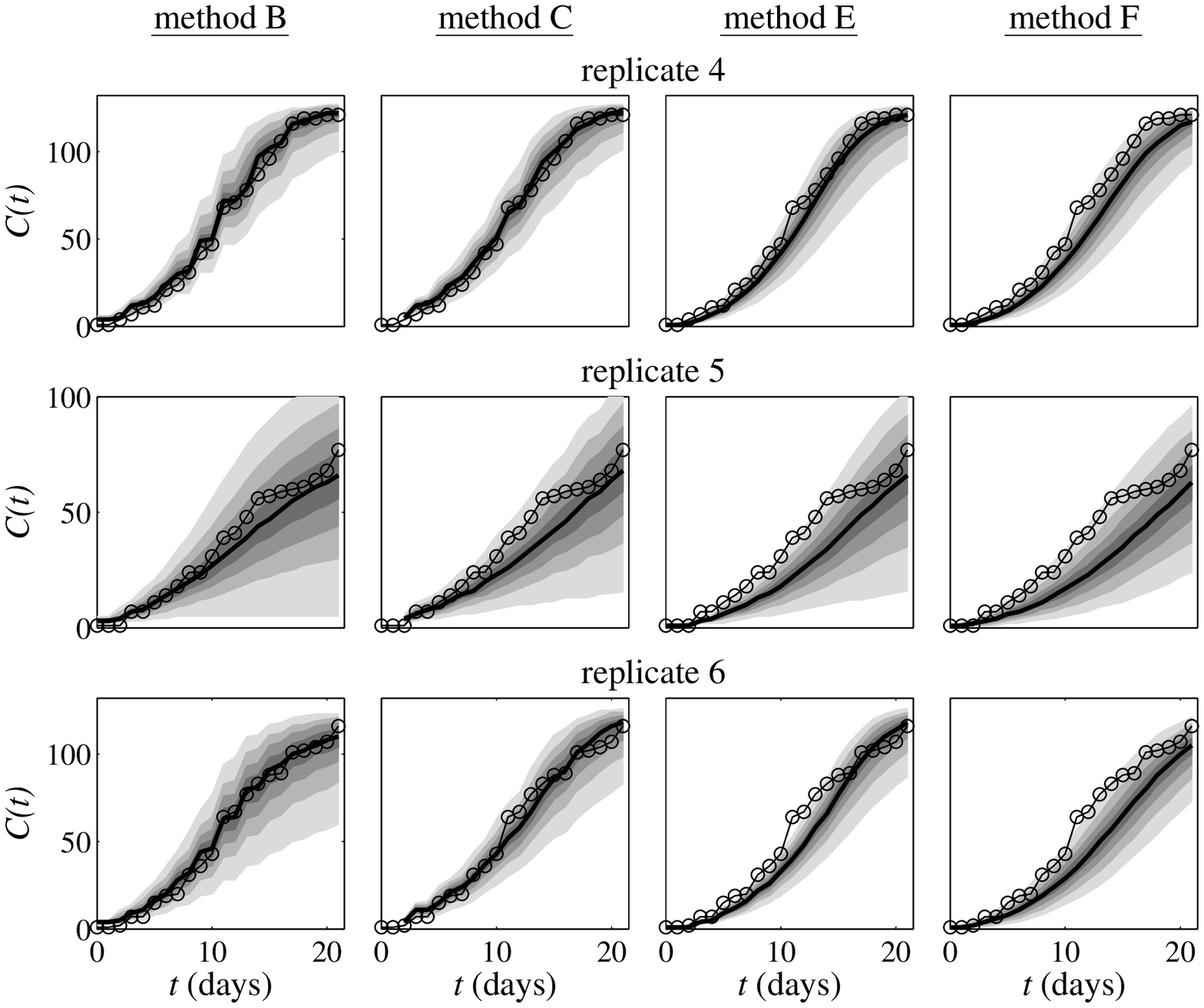}
\caption{\label{fig:DPC_10mm_2}
Same as in Fig.~S\ref{fig:DPC_10mm_1}, for replicates 4-6.}
\end{figure}

Figs.~S\ref{fig:DPC_10mm_1} and \ref{fig:DPC_10mm_2} show the comparison of the experimental incidence for systems with lattice spacing $a=10$~mm with the estimations $\rho(C|t)$ obtained from observations during time $t\leq t_{\text{obs}}=21$~days. Comparisons are presented for methodologies B, C, E, and F. 
Results for methodology A (based on the  MD method in step (iii) for prediction) are comparable to those obtained by methodology B (based on the ABC method) and have not been plotted in Figs.~S\ref{fig:DPC_10mm_1} and \ref{fig:DPC_10mm_2} for clarity. Similarly, the comparison for methodology D (based on the ABC inference method) is similar to that plotted for methodology C (based on the MD method) (not shown in Figs.~S\ref{fig:DPC_10mm_1} and \ref{fig:DPC_10mm_2}). These results imply that, given a level of description of data (step (i)) and model in step (ii), results are quite independent of whether MD or ABC approaches are used to address step (iii).

As can be seen from the figures, the p.d.f $\rho(C|t)$ obtained by methods C, D, and E give a better description of the observed data for most of the replicates.
On the other hand, the p.d.f. obtained by methodology F are often not able to capture the global trend of the incidence (see, e.g., replicate 3). 
This suggests that the focus of methodology F on finer details of the evolution may prevent the predictability of the invasive properties of the system.
As discussed in the main text, this might be due to the negative interplay between the simplicity of the CT model and the individual-based description of the data-augmented MCMC method for inference.

\section{Forecast of the incidence}
\label{sec:incidence_forecast}
In the main text, we quantified the differences between predictions and observations for all methods and fungal invasion experiments in terms of the quantities $\Delta c$ and $\Delta F$ (cf. Fig.~6 of the main text). In this section, we give a more visual comparison between observations and predictions based on the temporal incidence, $C(t)$.

\begin{figure}[h]
\begin{center}

\begin{pspicture}(0,0)(16,10)

\rput(8,5){\includegraphics[clip=true,width=16cm]{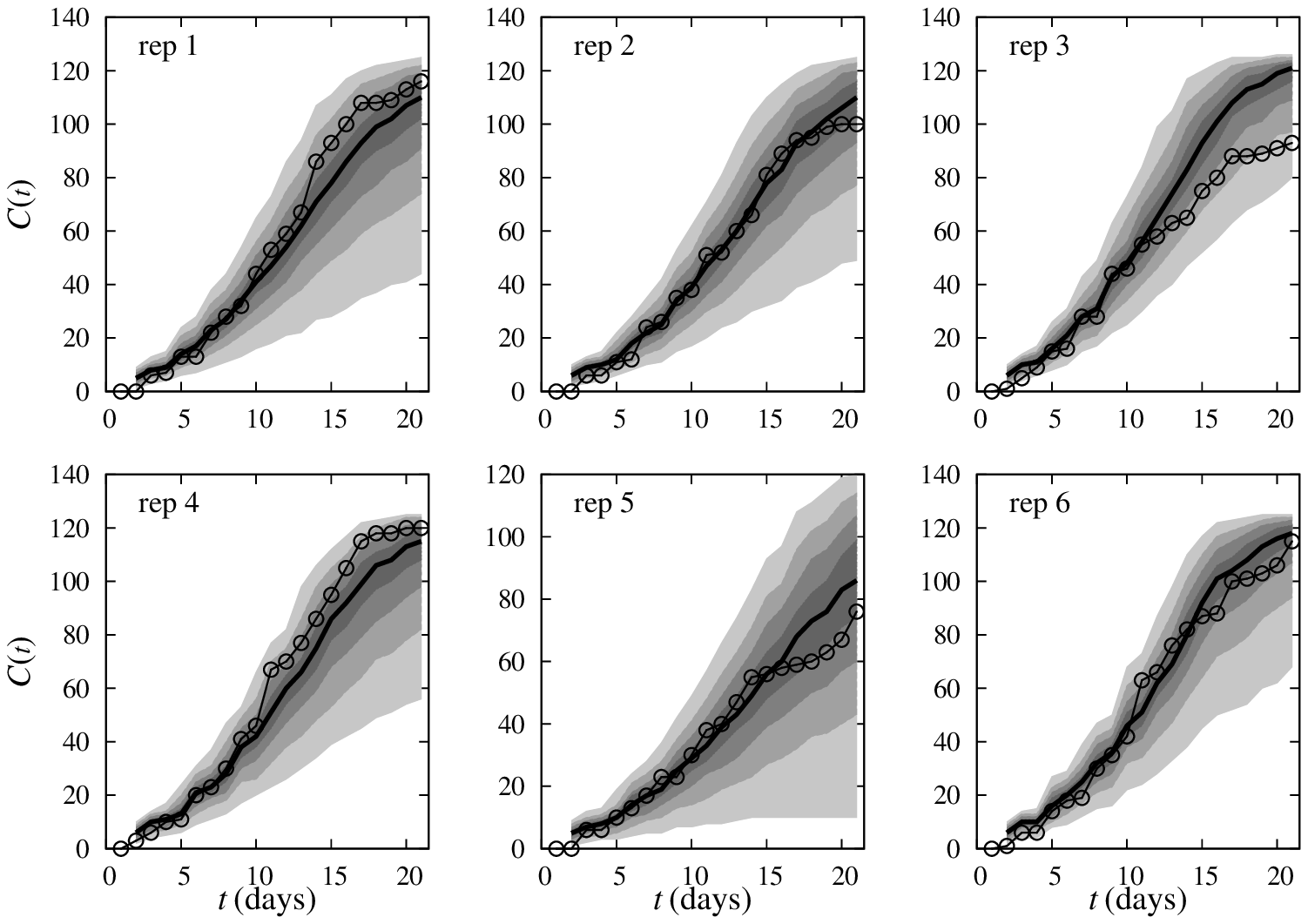}}
\psline[arrowsize=6pt]{->}(3.1,8.2)(3.1,7.5)
\psline[arrowsize=6pt]{->}(7.88,8.2)(7.88,7.5)
\psline[arrowsize=6pt]{->}(12.65,8.3)(12.65,7.6)

\psline[arrowsize=6pt]{->}(3.1,3.1)(3.1,2.4)
\psline[arrowsize=6pt]{->}(7.88,3.1)(7.88,2.4)
\psline[arrowsize=6pt]{->}(12.65,3.1)(12.65,2.4)

\end{pspicture}

\end{center}
\caption{ \label{fig:Fore_10mm_tobs_10} Forecast of the incidence for
  fungal invasion in a lattice of agar dots with $a=10$~mm. Each panel
  corresponds to a different experimental replicate of the
  epidemic. 
  The grey-scaled shaded
  area shows the p.d.f. $\rho(C|t)$ for the numerically
  extrapolated incidence based on observations
  over time $t \leq t_{\text{obs}}=10$~days, as marked by arrows. The
  grey-scale shaded areas are the $20\%$ (darker), $40\%$, $60\%$, and $80\%$
  (lighter) percentiles around the median (bold solid line).}
\end{figure}

\begin{figure}[h]
\begin{center}

\begin{pspicture}(0,0)(16,10)

\rput(8,5){\includegraphics[clip=true,width=16cm]{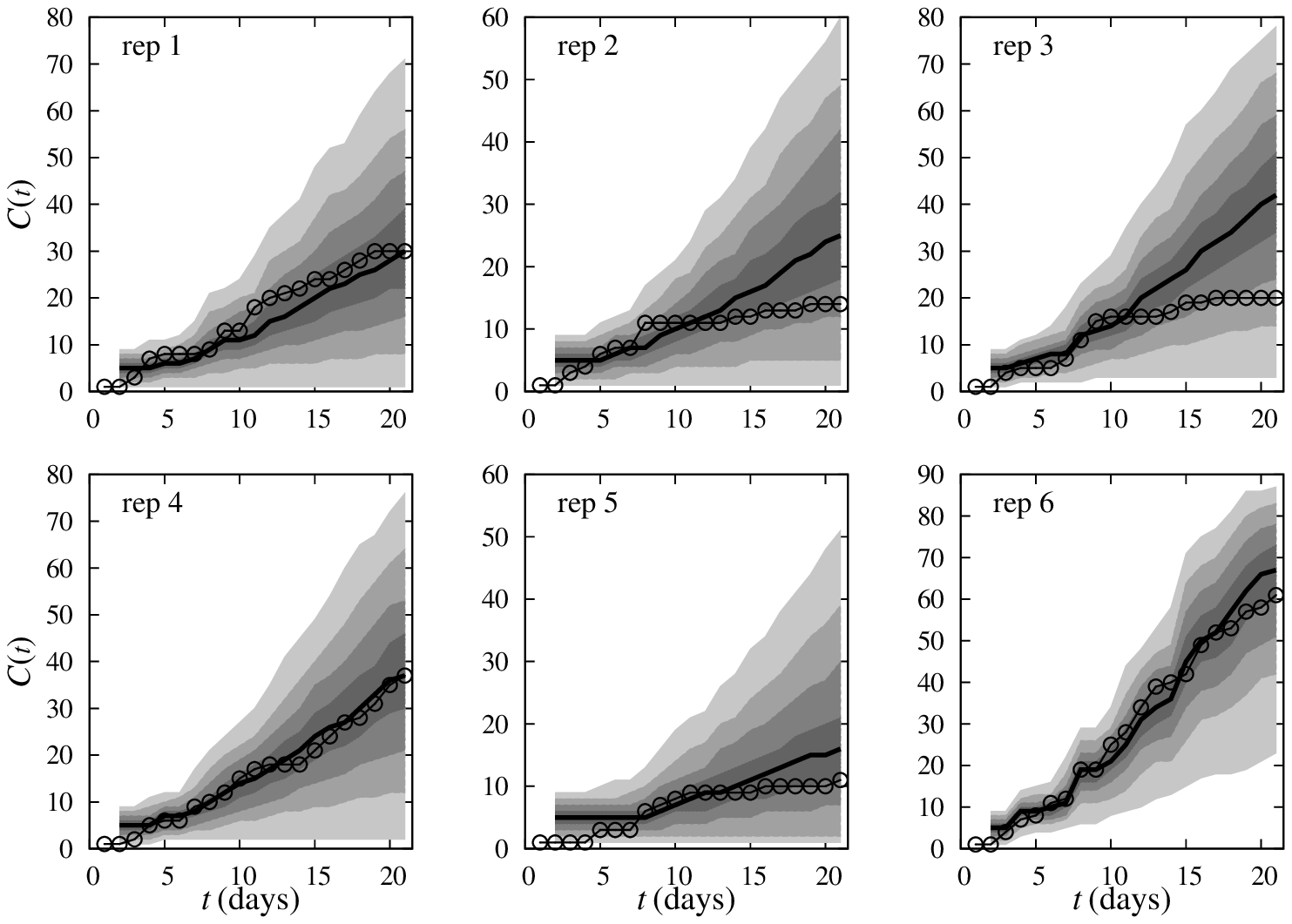}}
\psline[arrowsize=6pt]{->}(3.1,7.7)(3.1,7.0)
\psline[arrowsize=6pt]{->}(7.88,7.7)(7.88,7.0)
\psline[arrowsize=6pt]{->}(12.65,7.7)(12.65,7.0)

\psline[arrowsize=6pt]{->}(3.1,2.7)(3.1,2.0)
\psline[arrowsize=6pt]{->}(7.88,2.7)(7.88,2.0)
\psline[arrowsize=6pt]{->}(12.65,2.7)(12.65,2.0)
\end{pspicture}

\end{center}
\caption{ \label{fig:Fore_12mm_tobs_10} Similar representation as in
  Fig.~S\ref{fig:Fore_10mm_tobs_10} for fungal epidemics in
  populations of agar dots with lattice spacing $a=12$~mm.}

\end{figure}

Figs.~S\ref{fig:Fore_10mm_tobs_10} and S\ref{fig:Fore_12mm_tobs_10} show the comparison for the six
replicates available in the experiments with $a=10$~mm and $a=12$~mm,
respectively. Predictions of the incidence between days 11 and 21 are based on estimates of the transmission parameters made during the first 10 days with methodology C. As can be seen, the estimated incidence provides a
reasonable statistical description for the observed incidence for all
the replicates.  The quantity $\Delta c$ quantifies the rms distance between the observed incidence and the predicted incidence obeying the p.d.f $\rho(C|t)$ (given by the grey-scale shaded area in Figs.~S\ref{fig:Fore_10mm_tobs_10} and S\ref{fig:Fore_12mm_tobs_10}).

\section{Additional results for the numerical experiments}

\subsection{Homogeneous transmission of infection}
\label{sec:Homogeneous}
\begin{figure}[h]
\begin{center}
{\includegraphics[clip=true,width=12cm]{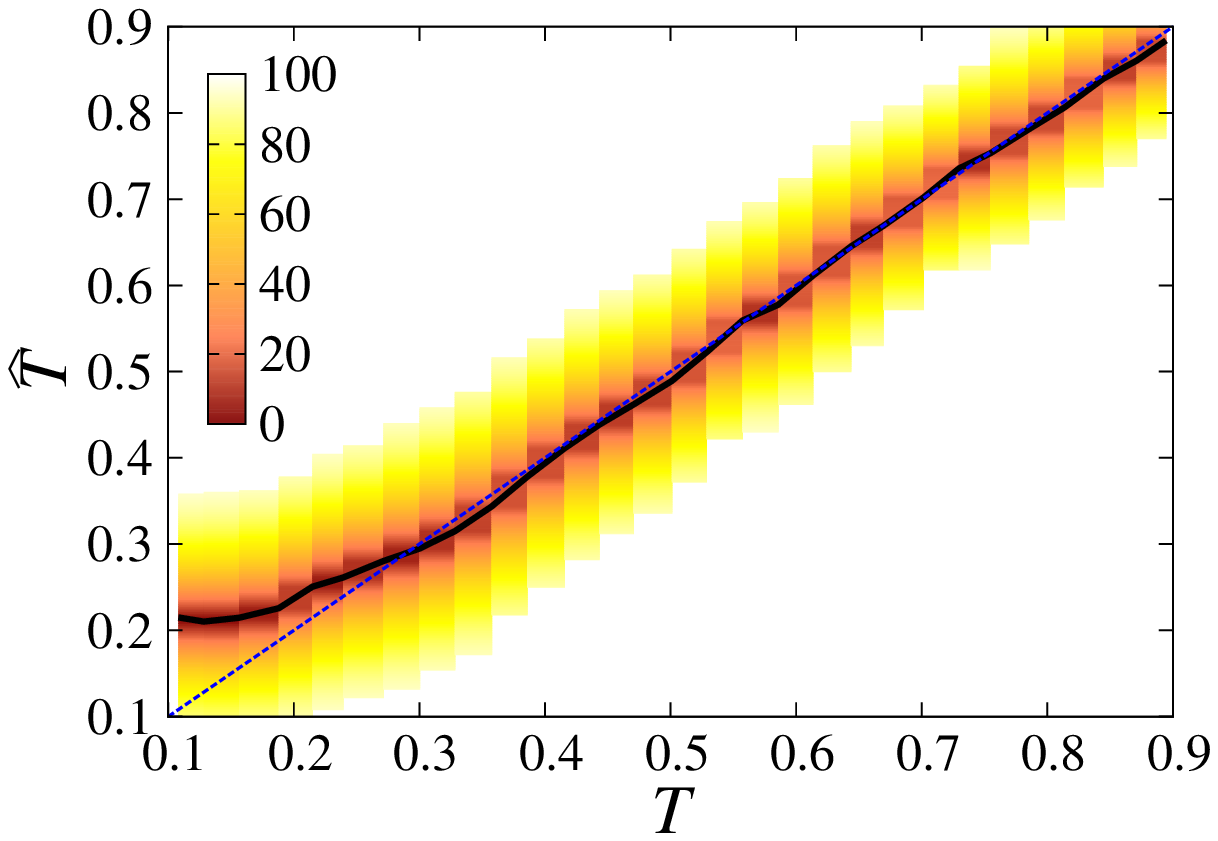}}
\end{center}
\vskip-10pt
\caption{ 
\label{fig:hatTvsT} 
Estimates of the transmissibility for numerical SIR epidemics with
homogeneous transmissibility. The shaded yellow-brown region shows the
levels of confidence as percentage of the
p.d.f. $\langle \rho(\hT|T) \rangle_{\text{e}}$ around the most probable
mean transmissibility, $\langle \hT_{\star} \rangle_{\text{e}}$ (continuous line), corresponding
to each value of $T$.  The dashed line representing the ideal situation (i.e. exact prediction)
with $\hat{T}=T$ 
is given for comparison with the actual prediction shown by the
  continuous line.
}
\end{figure}

Here, we give numerical support to the claim made in the main text
that the most probable estimate for transmissibility, $\hT_{\star}$,
corresponding to the maximum of the probability density function
(p.d.f.) $\rho(\hT)$, does not differ significantly from the actual
transmissibility, $T$.  
This is shown in Fig.~\ref{fig:hatTvsT} where
the estimate for the transmissibility, $\hT$,
is plotted as a function
of the transmissibility $T$. 
The estimates have been obtained for many different SIR numerical 
epidemics ($\sim 10^4$) with $T \in [0,1]$. 
The analysis has been restricted to epidemics
with final size $N_{\text{R}}$ (i.e. the number of removed hosts) greater
than a certain cut-off, $N_0$, in order to avoid estimates for small
epidemics giving a poor estimator for transmissibility. 
Excluding
small epidemics from the analysis also makes sense from a practical
point of view because they are not a threat in terms of invasion. 
We 
have checked that the statistics for $\hT$ do not depend
significantly on $N_0$ for $N_0 \gtrsim
5$.   
For each epidemic $i$ with given $T$, we obtain the
p.d.f. $\rho_i(\hT|T)$ for the effective transmissibility, $\hT$, from
observations of the initial stage (for $t
\le \tobs = 7\tau$, where $\tau$ is the infectious period for infected hosts which is taken as the unit of time, $\tau=1$). 
Then, the mean p.d.f.,
$\langle \rho(\hT|T) \rangle_{\text{e}}$, is calculated by averaging $\rho_i(\hT|T)$ over
$N_{\text{e}}(T)$ different stochastic realisations of epidemics with
given value of $T$:
\[
\langle \rho(\hT|T) \rangle_{\text{e}}=\frac{1}{N_{\text{e}}(T)}\sum_{i=1}^{N_{\text{e}}(T)} \rho_i(\hT|T)~.
\]
The first moment of $\langle \rho(\hT|T) \rangle_{\text{e}}$ gives an
estimate for the mean 
$\langle \hT_{\star} \rangle_{\text{e}}$ averaged over stochastic realisations. The dependence
of $\langle \hT_{\star} \rangle_{\text{e}}$ on $T$ is shown by the continuous line in
Fig.~S\ref{fig:hatTvsT}. The dispersion of $\langle \rho(\hT|T) \rangle_{\text{e}}$, shown by
the shaded region in Fig.~S\ref{fig:hatTvsT} as a function of $T$,
contains contributions from both the width of each individual
distribution, $\rho_i(\hT|T)$, and dispersion of the maxima for
  different replicates. In particular, the standard deviation, $\sigma(T)$,
of $\langle \rho(\hT|T) \rangle_{\text{e}}$ is given by
\[
\sigma(T)=\left[\langle \sigma_i^2 \rangle_{\text{e}}+\sigma_{\star}^2 \right]^{1/2}~,
\]
where
\[
\langle \sigma_i^2 \rangle_{\text{e}}=\frac{1}{N_{\text{e}}(T)} \sum_{i=1}^{N_{\text{e}}(T)}
\left[ \int_0^1 \hT^2 \rho_i(\hT|T) \text{d}\hT - \left( \int_0^1
    \hT \rho_i(\hT|T) \text{d}\hT \right)^2\right]
\]
is the average over stochastic realisations of the variance $\sigma_i^2$ of
$\rho_i(\hT|T)$. The quantity $\sigma_{\star}^2$ is the variance of
$\hT_{\star,i}$ over stochastic realisations, calculated as 
\[
\sigma_{\star}^2 = \frac{\sum_{i=1}^{N_{\text{e}}(T)} \hT_{\star,i}^2}{N_{\text{e}}(T)} 
-\left( \frac{\sum_{i=1}^{N_{\text{e}}(T)}
  \hT_{\star,i}}{N_{\text{e}}(T)}  \right)^2~.
\]

As can be seen from Fig.~S\ref{fig:hatTvsT} the actual value for the
transmissibility is statistically well described by the distribution
$\langle \rho(\hT|T) \rangle_{\text{e}}$. The mean for the most probable estimate for the transmissibility,
$\langle \hT_{\star} \rangle_{\text{e}}$, is in good agreement with the actual
transmissibility. In particular, $\langle \hT_{\star} \rangle_{\text{e}}$ provides an excellent
estimate for $T$ in the most interesting situations with $T \gtrsim
0.3$ where invasion is more likely. The deviations of $\langle
  \hT_{\star} \rangle_{\text{e}}$ from $T$ are larger for
small values of $T$ because epidemics are typically small and the deviations are large.
However, it is important to note that
$\langle \hT_{\star} \rangle_{\text{e}}$ overestimates $T$ in these
situations and thus provides a safe bound for
invasion.

\subsection{Heterogeneous transmission of infection} 
\label{sec:Heterogeneous}
In the main text, we have analysed the epidemics in model systems with
  homogeneous transmissibility for all pairs of connected hosts. 
Realistic populations of hosts exhibit inherent heterogeneity in
transmissibility and it is crucial to understand its effect on the prediction method introduced in the main
text.

In order to study the predictability of invasion for epidemics with heterogeneity in transmission of infection, we consider a simple but generic situation in which
transmission is heterogeneous due to variability  in the infectivity,
$\I$, and susceptibility, $\S$, of hosts.  
As a first approximation,
the rate of infection from an infected donor host with infectivity
$\I_{\text{d}}$ to a susceptible recipient host with susceptibility $\S_{\text{r}}$
is defined as $\beta_{\text{d-r}}=\I_{\text{d}}\S_{\text{r}}$
\cite{Miller_JApplProbab2008}.
We assume that $\I_{\text{d}}$ and $\S_{\text{r}}$ are independent random variables
distributed according to truncated normal distributions, ${\cal
  N}(\barI,\sigma_I^2)$ and ${\cal N}(\barS,\sigma_S^2)$, respectively,
which are the same for each host\footnote{The support of the normal
  distributions has been restricted to $[0,\infty)$ to ensure that
  both $\I_i$ and $\S_i$ are positive.}.
The mean values, $\barI$ and $\barS$, provide an effective measure of
the mean strength of the transmissibility while the standard
deviations, $\sigmaI$ and $\sigmaS$, characterize the degree of
heterogeneity. 
The multiplicative form of the infection transmission rate
$\beta_{\text{d-r}}=\I_{\text{d}}\S_{\text{r}}$ brings correlations in
transmissibilities, $T_{\text{d-r}}=1-e^{\tau
  \beta_{\text{d-r}}}$~\cite{grassberger1983}.  
Indeed, all the transmissibilities from a donor
are affected by the value of $\I_{\text{d}}$ and thus they are not
independent.  
Similarly, all the transmissibilities to a recipient are
influenced by its susceptibility $\S_{\text{r}}$ and thus also
correlated.  
Such correlations make the invasion probability for
heterogeneous system to be dependent on the whole set of
transmissibilities $\{T_{\text{d-r}}\}$.  
In spite of that, the method 
based on a single effective transmissibility, $T$,
 and Eq.~[1] in the main text  
is still applicable and useful. 
In realistic situations, the 
  transmissibility is often assumed to be homogeneous because the precise
degree of heterogeneity in transmission of infection is unknown and
difficult to infer in detail.

\begin{figure}[h]
\begin{center}
{\includegraphics[clip=true,width=12cm]{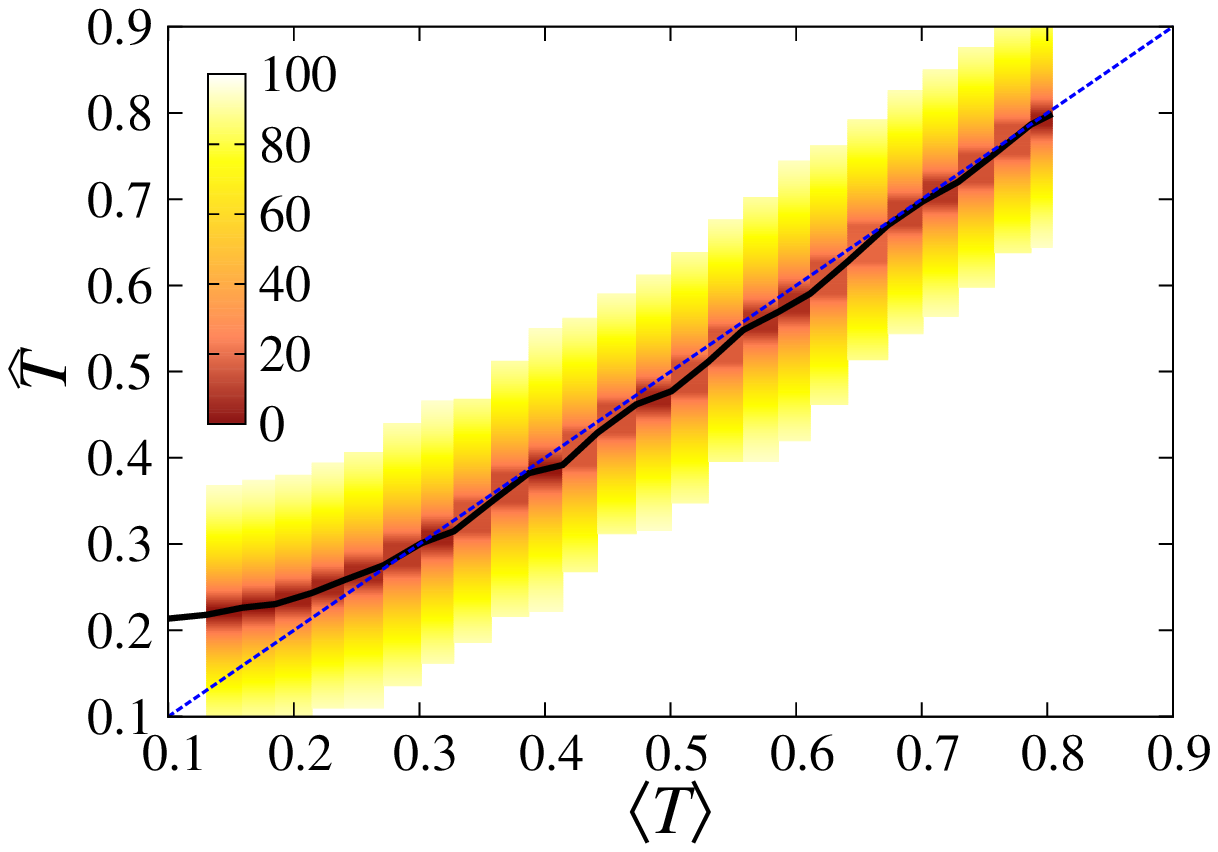}}
\end{center}
\caption{ \label{fig:hatTvsT_Het} Estimates for the effective transmissibility
  in numerical SIR epidemics with heterogeneous transmissibility
  introduced by Gaussian randomness in infectivity $\I$ and
  susceptibility $\S$ (with $\sigmaI=\sigmaS=0.2$, mean infectivity  
  set to $\barI=0.4$, and variable mean susceptibility $0.1 \le \barS
  \le 3.5$). The shaded yellow-brown region shows the
levels of confidence as percentage of the
p.d.f. $\langle \rho(\hT|\langle T \rangle) \rangle_{\text{e}}$ around the most probable
mean transmissibility, $\langle \hT_{\star} \rangle_{\text{e}}$ (continuous line).  
The dashed line representing the ideal situation (i.e. exact
   prediction) 
with $ \hT=\langle T \rangle$ 
is given for comparison with actual prediction shown by the
  continuous line.
}
\end{figure}

\begin{figure}
\begin{center}
{\includegraphics[clip=true,width=10cm]{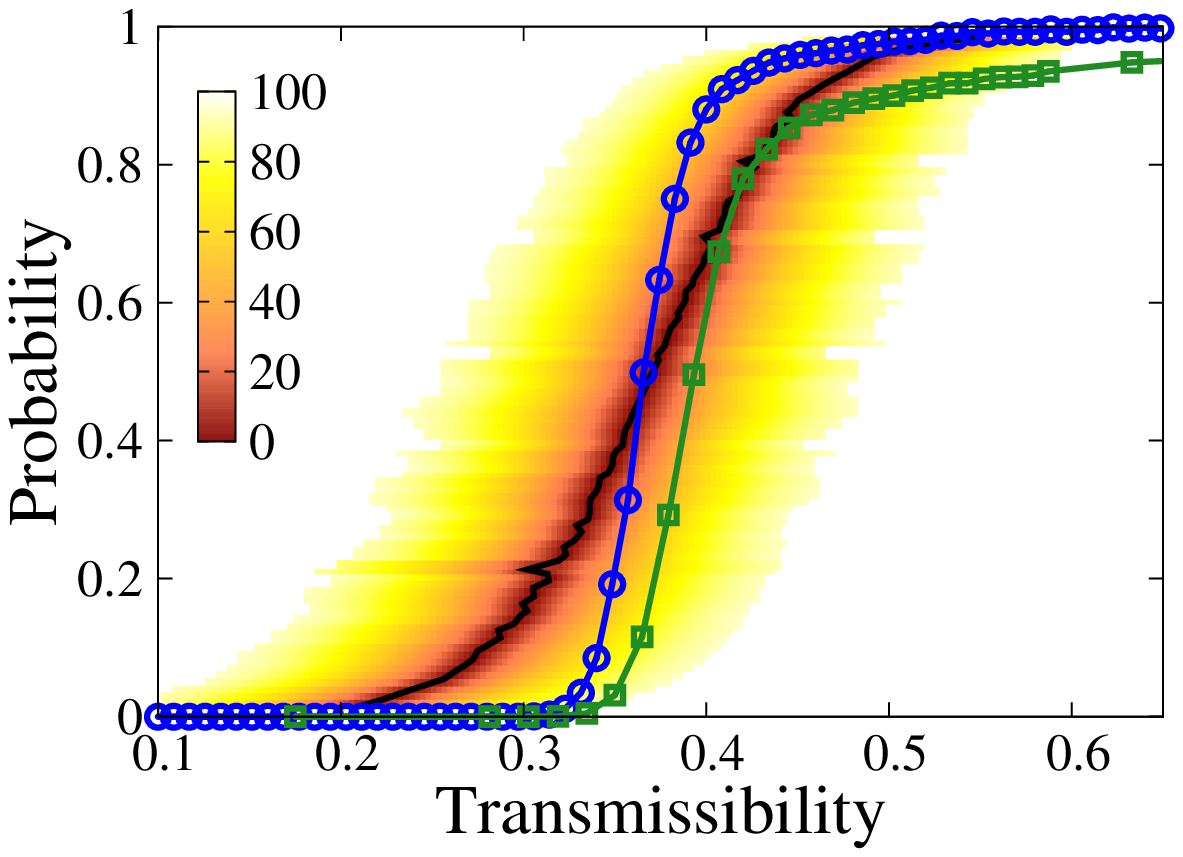}}
\end{center}
\caption{ \label{fig:PinvHet} Numerical experiments of SIR
    epidemics with heterogeneous transmissibility induced by Gaussian
    randomness in infectivity $\I$ and susceptibility $\S$ (with
    $\sigmaI=\sigmaS=0.2$, mean infectivity set to $\barI=0.4$, and
    variable mean susceptibility $0.1 \le \barS \le 3.5$).  The lines
    marked by circles and squares correspond to the probabilities of
    invasion $\Phom$ and $\Phet$ plotted \emph{vs} $\langle T
    \rangle$ for homogeneous and heterogeneous
    systems of size $L=51$, respectively,.
    Estimates for the probability of invasion, $\hPinv(L)$, have been
    evaluated for each epidemic (out of $\sim 10^4$) using the
    p.d.f. $\rho(\hT)$ obtained by observing the initial evolution
    during time $t \leq t_{\text{obs}}=7 \tau$ and then fitting the
    observed spatio-temporal map by the shell-evolution function.  
    Horizontal slices of the yellow-brown
    shaded area corresponding to a fixed value of $\hPinv(L)$
    represent the distribution $\rho(\hT)$ averaged over realisations
    of epidemics with the same value of $\hPinv(L)$.  
    The points on the ridge (black solid curve) of the shaded area
    correspond to the 
    most probable transmissibility, $\hT_*$, averaged over epidemics with a
    certain value of $\hPinv(L)$.  }
\end{figure}

In order to test our methodology in heterogeneous systems with different
strengths of transmissibility, we perform  numerical experiments
for epidemics with $\sigmaS = 
\sigmaI = 0.2$, mean infectivity set to $\barI=0.4$ and variable
$\barS$. 
Again, observations are made over an
initial interval of time 
$t \le t_{\text{obs}} = 7 \tau$ 
for estimation of the effective transmissibility, $\hT$. 
Similarly to the analysis given in the previous section for
epidemics with homogeneous transmission, Fig.~S\ref{fig:hatTvsT_Het} shows a comparison between the
  p.d.f. $\langle \rho(\hT|\langle T \rangle) \rangle_{\text{e}}$
averaged over stochastic realisations of epidemics (shaded region)
with the  ideal situation giving exact prediction of the
  spatially averaged 
  transmissibility used in the simulations, i.e. $\hat{T}=\langle T
  \rangle$ (dashed line). 
As can be seen, the estimates are statistically
consistent with $\langle T \rangle$. In fact, the mean of the most
probable transmissibility, $\langle \hT_{\star} \rangle_{\text{e}}$,
gives a good description for $\langle T \rangle$ (compare the
continuous and dashed lines in Fig.~S\ref{fig:hatTvsT_Het}).
We proceed further
as in the case with homogeneous transmission
by calculating $\hPinv(L)$ for each of the numerical epidemics 
in a system of size $L=51$
(see Fig.~2(a) in the main text) by using Eq.~[1] in the main text, the
estimated $\rho(\hT)$, and the probability of invasion $\Phom(\hT;L)$ for
systems with homogeneous transmissibility equal to $\hT$. 
Note that here $\Phom(\hT;L)$
corresponds to the function denoted as $\Pinv(\hT;L)$ in the main
text.
The notations have been changed in order to distinguish between the probability
of invasion in homogeneous systems and the probability of invasion in
the presence of heterogeneity, denoted as $\Phet(\hT;L)$.
The results of our estimations are shown in
Fig.~S\ref{fig:PinvHet}. 
The relation between $\hPinv(L)$ and
$\Phom(\hT;L)$ is very similar to the relation reported in the main
text for epidemics with homogeneous transmission. 
Indeed, for
epidemics with low transmissibility, $\hPinv(L)$ typically
overestimates $\Pinv$. 
In contrast, for more invasive epidemics,
$\hPinv(L)$ underestimates $\Pinv$ for most of the possible effective
transmissibilities. 
Although this comparison has some interest, in the current
situation it makes more sense to compare the estimated probability of
invasion with the actual
probability of invasion in heterogeneous system, $\Phet(\hT;L)$, that can be calculated
numerically and shown by the line with squares in
Fig.~S\ref{fig:PinvHet}. 
The comparison of this line with the shaded
region reveals that the estimated $\hPinv(L)$ overestimates the actual
probability of invasion in most of the situations both for cases with
low and high transmissibility. 
Therefore, the  estimates of $\hPinv$ typically provide
safe bounds for the probability of invasion. 
In fact, the
larger the heterogeneity in susceptibility and/or infectivity, the
safer the bound is. 
This is a consequence of the inequality
$\Pinv(\langle T \rangle) \ge \Phet(\langle T \rangle)$ that holds for
any given value of $\langle T \rangle$ under quite general conditions
due to the existence of correlations in transmission induced by
heterogeneity in the transmission rates 
\cite{Cox1988,Miller_JApplProbab2008}. 
Indeed, Fig.~S\ref{fig:PinvHet}
 shows that the inequality
holds for the numerical experiments considered here with $\langle T \rangle=\hT$
(i.e. the line corresponding to the heterogeneous case marked by the
  squares is below the line marked by the circles for homogeneous
  system). 
These results are
particularly encouraging for analysis of realistic epidemics in which
a certain degree of heterogeneity in susceptibility and infectivity of
hosts is expected to be ubiquitous.

\section{Differences between $P_{\text{inv}}$ and $\hat{P}_{\text{inv}}$}
\label{sec:Pinv}
In this section, we discuss the origin of the difference between the estimated probability of
invasion, $\hat{P}_{\text{inv}}(L)$, evaluated at the most probable
transmissibility, $\hT_{\star}$, and the actual probability of
invasion $\Pinv(\hT;L)$ one would obtain if the transmissibility of
the epidemic was known exactly. 

To start with, recall that the relation between
$\hat{P}_{\text{inv}}(L)$ and $\Pinv(\hT;L)$ is given by Eq.~[1] in
the main text. 
As we have
seen, $\rho(\hT)$ is a peak-shaped function approximately
symmetric about the peak position at $\hT_{\star}$.
Therefore, the peak region  will mainly
  contribute to the integral in Eq.~[1] of the main text.
If $\hT_{\star}$ is not too close to the inflection
point of $\Pinv(\hT;L)$, 
the Taylor series expansion of $\Pinv(\hT;L)$ in $(\hT
  -\hT_{\star})$ to second  order, i.e., 
\begin{equation}
\label{eq:PinvExpansion}
\Pinv(\hT;L) = \Pinv(\hT_{\star};L)+ \Pinv^{\prime}(\hT_{\star};L) (\hT-\hT_{\star}) +
\frac{1}{2} \Pinv^{\prime \prime}(\hT_{\star};L) (\hT-\hT_{\star})^2,
\tag{S.10}
\end{equation}
is sufficient to estimate  the deviation of $\hPinv$ from $\Pinv$. 
Indeed, substitution of Eq.~\eqref{eq:PinvExpansion} into Eq.~[1] of the main text gives:
\begin{equation}
\label{eq:hPinvExpansion}
\hPinv(L)=\Pinv(\hT_{\star};L)+ \Pinv^{\prime}(\hT_{\star};L) \int_0^1 \rho(\hT)(\hT-\hT_{\star}) \text{d}\hT  +
\frac{1}{2} \Pinv^{\prime \prime}(\hT_{\star};L) \int_0^1 \rho(\hT)
(\hT-\hT_{\star})^2 \text{d}\hT
\tag{S.11}
\end{equation}

The distribution of $\hT$ is approximately symmetric around the
maximum, i.e. $\rho(\hT-\hT_{\star})\simeq \rho(\hT_{\star}-\hT)$ 
(see Fig.~2(b) in the main text), and thus the term containing
$\Pinv^{\prime}$ in Eq.~\eqref{eq:hPinvExpansion} is negligible
in comparison with other terms in the sum. 
This means that 
\begin{align}
\hPinv(L)>\Pinv(\hT_{\star};L) \text{ if }  \Pinv^{\prime
  \prime}(\hT_{\star};L)>0
\nonumber
\\
\hPinv(L) \leq \Pinv(\hT_{\star};L) \text{ if }  \Pinv^{\prime
  \prime}(\hT_{\star};L) \leq 0~,
\tag{S.12}
\end{align}
where we have taken into account that the last integral in
  Eq.~\eqref{eq:hPinvExpansion} is positive. 
The above inequalities demonstrate that  the value and sign of
$\hPinv(L)- \Pinv(\hT_{\star};L)$ depends on the curvature of
$\Pinv(\hT;L)$ around $\hT=\hT_{\star}$ which is given by $\Pinv^{\prime \prime}$.



%